\documentclass[useAMS,usenatbib]{mn2e}
\usepackage{epsfig,aas_macros,amsmath}

\newif\ifAMStwofonts

\title[Rotation in Hydra-A]{Cold gas dynamics in Hydra-A: evidence for a rotating disc}

\author[Hamer et al.]
{\parbox[h]{\textwidth}
{S. L. Hamer,$^{1,6}$\thanks{E-mail: Stephen.Hamer@obspm.fr} A. C. Edge,$^{1}$ A. M. Swinbank,$^{1}$
J. B. R. Oonk,$^{2}$ R. Mittal,$^{3}$ \\
B. R. McNamara,$^{4}$  H. R. Russell,$^{4}$ M. N. Bremer,$^5$ F. Combes,$^6$ A. C. Fabian.$^7$ N. P. H. Nesvadba,$^8$ C. P. O'Dea,$^{12}$ S. A. Baum,$^3$ P. Salom\'e,$^6$ G. Tremblay,$^9$ M. Donahue,$^{10}$ G. J. Ferland,$^{11}$ \& C. L. Sarazin$^{13}$} \\
\vspace*{6pt}\\
$^{1}$Institute for Computational Cosmology, Department of Physics, Durham University, South Road, Durham DH1 3LE\\
$^{2}$ASTRON, Netherlands Institute for Radio Astronomy, PO Box 2, 7990 AA Dwingeloo, the Netherlands\\
$^{3}$Chester F. Carlson Center for Imaging Science, 54 Lomb Memorial Drive, Rochester, NY 14523, USA\\
$^{4}$Department of Physics and Astronomy, University of Waterloo, Waterloo, ON N2L 3G1, Canada \\
$^{5}$H H Wills Physics Laboratory, Tyndall Avenue, Bristol BS8 1TL\\
$^{6}$LERMA Observatoire de paris, CNRS, 61 rue de l'Observatoire, 75014 Paris, France\\
$^{7}$Institute of Astronomy, University of Cambridge, Madingley Road, Cambridge, CB1 0HA\\
$^{8}$Institut d'Astrophysique Spatiale, CNRS, Universit\'e Paris-Sud, 91405 Orsay, France\\
$^{9}$European Southern Observatory, Karl-Schwarzschild-Str. 2, 85748 Garching bei M\"{u}nchen, Germany \\
$^{10}$Physics and Astronomy Department, Michigan State University, East Lansing, MI 48824, USA \\
$^{11}$Department of Physics, University of Kentucky, Lexington, KY 40506, USA \\
$^{12}$School of Physics and Astronomy, RIT, 84 Lomb Memorial Dr., Rochester, NY 14623 \\
$^{13}$Department of Astronomy, University of Virginia, Charlottesville, VA 22904-4325 \\
}

\begin{document}

\date{Accepted 09 October 2013. Received ; in original form }

\pagerange{\pageref{firstpage}--\pageref{lastpage}} \pubyear{2013}

\maketitle

\label{firstpage}

\begin{abstract}

\noindent We present multi-frequency observations of the radio galaxy 
Hydra-A (3C218) located in the core of a massive, X-ray luminous galaxy cluster.
IFU spectroscopy is used to trace the kinematics of the ionised and warm molecular 
hydrogen which are consistent with a $\sim$ 5~kpc rotating disc.
Broad, double-peaked lines of CO(2-1), [CII]157$\mu$m and [OI]63$\mu$m are detected.
We estimate the mass of the cold gas within the 
disc to be M$_{gas}$ = 2.3 $\pm$ 0.3 $\times$ 10$^9$ M$_{\odot}$.
These observations demonstrate that the complex line profiles found in the cold atomic
and molecular gas are related to the rotating disc or ring of gas.
Finally, an {\em HST} image of the galaxy shows that this gas disc contains a 
substantial mass of dust. 
The large gas mass, SFR and kinematics are consistent with the levels of 
gas cooling from the ICM.
We conclude that the cold gas originates from the continual quiescent 
accumulation of cooled ICM gas.
The rotation is in a plane perpendicular to the projected orientation 
of the radio jets and ICM cavities hinting at a possible 
connection between the kpc--scale cooling gas and the accretion of material 
onto the black hole.  We discuss the implications of these observations for 
models of cold accretion, AGN feedback and cooling flows.

\end{abstract}


\section{Introduction}

The symbiotic relationship between the brightest cluster galaxy (BCG) and the 
dense intracluster medium (ICM) that surrounds it has important implications
for our understanding of the formation and evolution of massive galaxies
within the cores of clusters \citep[see][for reviews]{fab12,mn12}.
The energetics of the gas show that the different gas phases are connected,
for instance the close correlation between the X-ray properties 
(cooling time, central entropy etc.)
of the cluster core and the detection of optical lines \citep{cra99} in the BCG
as shown by \citet{cav08} and \citet{san09}.
Hence, the implicit connection of the optical emission to the cold \citep{edg01,sc03} 
and warm \citep{edg02,ega06,don11} molecular gas,
suggests that there is a direct link between the rapidly cooling gas in the ICM
(t$_{\rm cool}$ $<10^9$~years) and the presence of cold molecular gas. 
However, any model positing that the cool gas in the BCG condenses from the 
hot ICM must also satisfy stringent observational constraints on the rate of 
energy loss from gas at temperatures above $\sim \frac{1}{3}$ T$_{\rm cluster}$
and strong limits on cooling of gas below that temperature
\citep{pet01,snd10}.  AGN in the BCG may supply the 
heat required to curtail radiative cooling. While the global AGN energy 
budget seems sufficient, the details of how the AGN prevents the ICM from rapidly 
cooling are not yet settled \citep{mn12,fab12}. 

Feedback of energy from AGN outbursts has been invoked both to provide a potential 
source of the heating required to truncate the cooling of the ICM and to resolve the 
over-production of massive
galaxies in semi-analytic models \citep{bow06,cro06}.
This AGN feedback injects mechanical 
energy into the ICM as outflows from the black hole inflate kpc--scale cavities in the 
X-ray  atmosphere \citep{mcn07}.
One of the most striking examples of these cavities is Hydra-A \citep[3C218,][]{mcn00,wis07}, 
which has clearly defined X-ray surface 
brightness depressions in the ICM that spatially correlate to 
the complex radio structure in this radio source. 
Hydra-A is the central galaxy of an X-ray luminous cluster which
exhibits strong optical line emission which
shows a complex velocity structure that clearly indicates evidence of rotation
\citep{sim79,es83} and dust \citep{han95}.
\citet{mcn95} identified the presence of excess blue light from
the central $\sim8''\times6''$ (8.4$\times$6.3 kpc) region which is spatially 
coincident 
with the rotating emission line nebula. This region
has a major axis which in projection is aligned almost perpendicularly to the radio
jet axis and suggests the presence of a young
stellar population within the rotating ionised gas which they 
attribute to locally enhanced star formation 
fuelled by either the cooling flow or in--falling material. \citet{mcn95} also 
noted that this configuration is similar to the standard model 
of AGN disc accretion, although the scales involved in this system are
much greater. 

Hydra-A shares many properties with the archetypal central galaxy in a cooling
flow cluster, NGC1275/Perseus-A \citep{fab81}, which is a powerful
radio galaxy with multiple cavities created by repeated AGN outbursts \citep{boh93,fab03}.
The detection of HI absorption \citep{dwa94,dwa95,tay96}, CO line
emission \citep{sc03}, PAH features \citep{don11} 
and warm H$_2$ ro-vibrational
and rotational lines \citep{jaf97,edg02,don11} in both Hydra-A and NGC1275 
all suggest a significant 
mass of atomic and molecular gas at a range of temperatures in
the core of both galaxies. The dynamics of this phase and its
relationship to the optical lines and associated star formation are open
issues given that most observations are from a single dish or single line of sight.

In this paper we present new data over sub-mm to optical wavelengths
that shed new light on the nature and dynamics of the cold gas in Hydra-A.
In particular, new data from {\sl Herschel} 
\citep{pil10} and
IRAM 30m combined with Very Large Telescope (VLT) integral field data provide a
much clearer picture of the dynamics of the cold gas.
In $\S$\ref{sec:obs} we outline our observations and analysis techniques.
Next we outline our results ($\S$\ref{sec:res}) before discussing their 
implications in $\S$\ref{sec:dis}.  Finally a summary and conclusions are 
reported in $\S$\ref{sec:sum5}.
We assume $\Omega_{m}=$ 0.27, $\Lambda=$ 0.73 and $H_{o}=$ 71 km s$^{-1}$ Mpc$^{-1}$ 
throughout. In this cosmology, The luminosity distance of the source is 242 Mpc and 
1$''$ corresponds to a physical scale of 1.053 kpc at 
the redshift of Hydra-A 0.054878 \citep{smi04}.

\section{Observations and Data Reduction}
\label{sec:obs}
\subsection{VIMOS}

Optical integral field unit (IFU) observations of Hydra-A were taken using the VIMOS (Visible 
Multiobject Spectrograph) instrument on the 8.2m Very Large Telescope (VLT) in October 
2007 (programme ID 080.A-0224(A)). 
Three 600 second exposures were taken with a 
pointing dither included between each exposure to minimise the effect of bad pixels. 
The HR$\_$Orange Grism and GG435 filter (spectral resolution of R\,$\sim$\,$\Delta\lambda / \lambda$\,$\sim$\,2650 over the wavelength range 5250--7400~\AA) were 
used to observe H$\alpha$ ($\lambda_{\mathrm rest} = 6562.8$~\AA) at the redshift of the 
cluster. The observations were taken in moderate seeing conditions with a mean 
seeing of $\sim$1.25$''$. 

We reduced the raw data using the {\sc esorex} package. 
We performed the basic data reduction 
including bias subtractions, flat fielding and the wavelength and flux 
calibrations.  In order to prepare the data for sky subtraction we masked any 
point--like objects to 
remove any stars within the field.  The BCG was then removed from the field 
by masking all pixels within an isophote of half the peak flux.
This level was chosen empirically as the best compromise to allow a 
good sampling of sky pixels while removing the majority of the BCG light.
The sky level for each quadrant is then the median 
value of the remaining pixels at each wavelength increment.  This sky 
spectrum was then subtracted from each pixel in the four quadrants before 
they were combined into a single datacube. The more extended galaxy contribution 
which was included in the sky spectrum is from stellar continuum and thus 
relativly flat. As our analysis is concentrated solely on the line emission this slight 
over-subtraction has a minimal effect on the results presented here.
Finally, we median--combined the 
three exposures for each pointing in order to eliminate cosmic rays.
To mosaic the three pointings we determined the exact position of the
BCG (the peak of the smoothed continuum flux) in each observation and combine 
them centring on this position to
create a cube with a $\sim$\,29$''$\,$\times$\,28$''$ field of view.


\subsection{SINFONI}

The SINFONI (Spectrograph for Integral Field Observations in the Near Infrared) 
instrument on the VLT was used to take H and K band IFU observations of Hydra-A
(programme 082.B-0671, PI Nesvadba) in November 2008 and February 2009.
The observations consist of three visits in each band, with each visit
comprising  six 300 second exposures, four centred on the BCG and two on
an offset sky region.
The observing conditions were better than those of the  
VIMOS observations with a mean seeing of $\sim0.59''$. 

The {\sc esorex} package was set up for SINFONI IFU observations and used to 
reduce these data. We corrected for dark current, the linearity 
of the detector, optical distortions between the slitlets and 
performed the basic calibration steps as for VIMOS. The sky reduction was also 
performed using this package for the separate sky observations taken during each 
pointing. The individual pointings in each band were then combined into a single 
cube in the same manner as the VIMOS observations. This procedure produced cubes with a 
$\sim$\,8$''$\,$\times$\,8$''$ field of view.

\subsection{Analysis of the IFU data}

The spectral cubes produced by the reduction processes were unwrapped into two 
dimensional spectral images so that the presence of 
emission lines could be identified by eye.  
Spectral models consisting of Gaussian emission lines and a flat continuum were fitted to each 
line at each spatial position so 
that maps of the spectral properties could be produced. The fitting was performed 
on a small region of the spectrum around the line to enable the fitting routine to 
work more effectively.  This region corresponded to 160 \AA \ (240 wavelength elements) 
for VIMOS and 490 \AA \ (200 wavelength elements) for SINFONI.  The noise level 
was calculated using a second such region near the first that contained no emission 
lines or sky features.  

The median value of the spectrum within the extracted region 
was taken as a continuum model against which to compare a fit to the line. 
The fits to the line were made using Gaussian models in which the parameters were 
allowed to vary. These fits were accepted when they provided a 7$\sigma$ 
better fit than the flat continuum alone. The primary fit to the VIMOS data was to 
the H$\alpha$/[NII] triplet.  The models were fit to the rest wavelength of the line 
and the redshift was allowed to vary, the three lines relative rest frame positions were 
fixed.  The total flux in the H$\alpha$ line and strong [NII]$\lambda_{rest}$ 6583 \AA~ line 
were allowed to vary independently, the flux in the weaker [NII] line was set at 
one third the flux of the brighter line.  The line widths (FWHM, deconvolved for instrumental 
resolution) were allowed to vary 
between pixels but the [NII] was forced to have the same width as the H$\alpha$.
The continuum was typically ``flat'' over the short wavelength ranges being studied.
As such the continuum level was set to a constant value throughout each spectral model.  
However, the value was allowed to vary between spatial resolution elements.  In 
lower flux regions where a 7$\sigma$ fit could not be obtained the routine binned 
the surrounding 8 pixels and attempted the fit again. Such binning may combine regions 
with different central velocities which can artificially broaden 
the lines.  This effect will be largest in regions with a high velocity gradient, we 
highlight cases where this may be a problem in the text.

In the SINFONI observations 
the brightest lines are from Paschen~$\alpha$ (Pa$\alpha$) and molecular hydrogen. The rest frame 
wavelengths of these lines are sufficiently separated to allow each line to be fitted 
independently from the other lines.
A single Gaussian model was fitted to the line at the rest wavelength 
and the redshift, flux, width (FWHM, deconvolved for instrumental resolution) and continuum level were allowed to vary.  Again the 
routine binned pixels together in the event a 7$\sigma$ fit could not be obtained.
When a line was detected in this way the single Gaussian model (or the triplet in the 
case of H$\alpha$ and [NII]) was tested against a two component model to check for 
extra velocity components. This second Gaussian was varied independently 
of the other model (central redshift, intensity and line width) and was allowed to be 
much broader.  When all the parameters were re-minimised, the $\chi^2$ of the new fit 
was 
calculated and compared to the original models. Although this provided a better fit 
to the data the extra components were found to not be significant due to the reduction 
of the degrees of freedom.

\subsection{IRAM 30m}

We obtained IRAM 30m data for the CO(1-0) and CO(2-1) lines in Hydra-A in April 
2010 at 109.274 and 218.545 GHz respectively. The observations were performed 
simultaneously in exceptional conditions
($\tau_{230\rm GHz}\sim 0.05$) with the EMIR receiver using a 4~GHz bandwidth 
covering each line \citep{car12}.  The observation was of 2 hours duration with wobbler 
switching with a 90$''$ throw.  The observations reached a noise level of 0.35~mK
and 0.30~mK in 100 km\,s$^{-1}$ channels for CO(1-0) and CO(2-1) respectively.
The unusual situation of obtaining a better temperature sensitivity at
CO(2-1), combined with a beam of a quarter the size compared to CO(1-0), means
that there is a highly significant CO(2-1) detection but at best a marginal
one for CO(1-0). 
Both the WILMA and 4~MHz backends were used to sample the 
data, but in this paper we present only the CO(2-1) spectrum taken with the WILMA
backend as they were the most stable and provide the most reliable CO detection.


\subsection{Herschel}

Hydra-A was observed with {\sl Herschel} as part of an Open Time Key Project
to observe a statistical sample of cooling flow clusters (programme OTKP$\_$aedge1, PI Edge).
We used the PACS instrument to obtain spectral observations of the two principal atomic 
cooling lines,
[OI] (OD:536, OBSID:1342207793) and [CII] (OD:538, OBSID:1342207817) at 63 and 
157~$\mu$m respectively. The lines were observed in line spectroscopy mode using the 
chopping and nodding implementation to account for the background and dark current.
These were then analysed using the standard pipeline routines included in the 
latest release of the Herschel Interactive Processing Environment (HIPE {\it v. 7.1.0}). 
A detailed description of the Herschel data reduction can be found in \citet{mit11}. 
The full analysis of these spectral data and the additional PACS and SPIRE photometry
will be made elsewhere (Oonk et al., in preparation).

\subsection{HST}

An {\sl HST} ACS image of Hydra-A was obtained on the 16th April 2011 using 
ACS with the WFC detector and 
F814W filter for an exposure time of 2367s. A 2709s UV image was obtained with {\sl HST} 
ACS on the 11th April 2011 using the SBC detector and the F140LP filter.
Standard calibrations were then 
applied including subtraction of the bias, correcting for dark current and flat 
fielding.
This observation was taken as 
part of a programme to obtain high resolution imaging of all the {\sl Herschel}
OTKP clusters (Proposal ID 12220, PI Mittal).

\section{Results}
\label{sec:res}

\subsection{IFU Maps}
\label{sec:maps}

\begin{figure*}
\psfig{figure=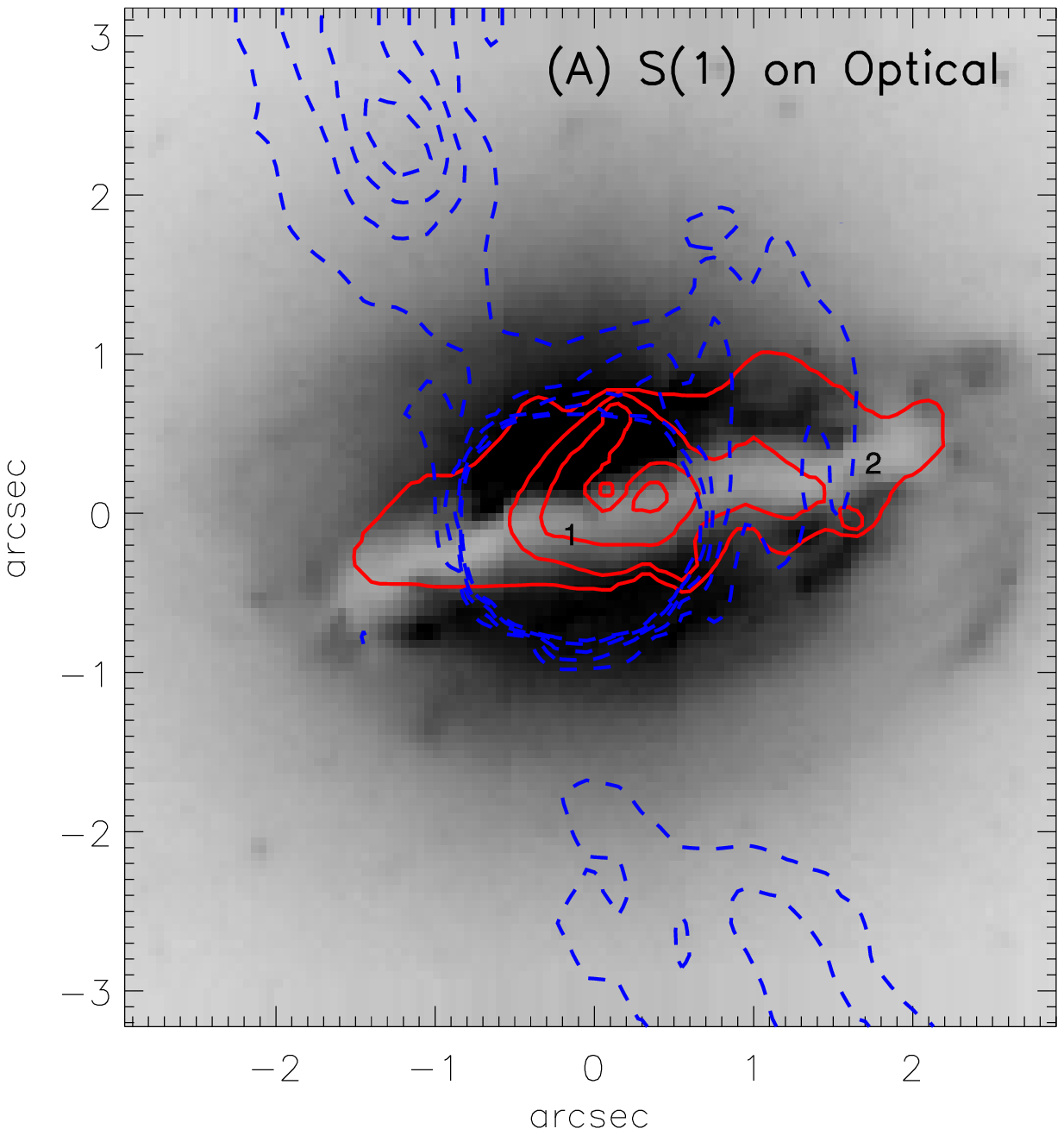,height=8cm}
\psfig{figure=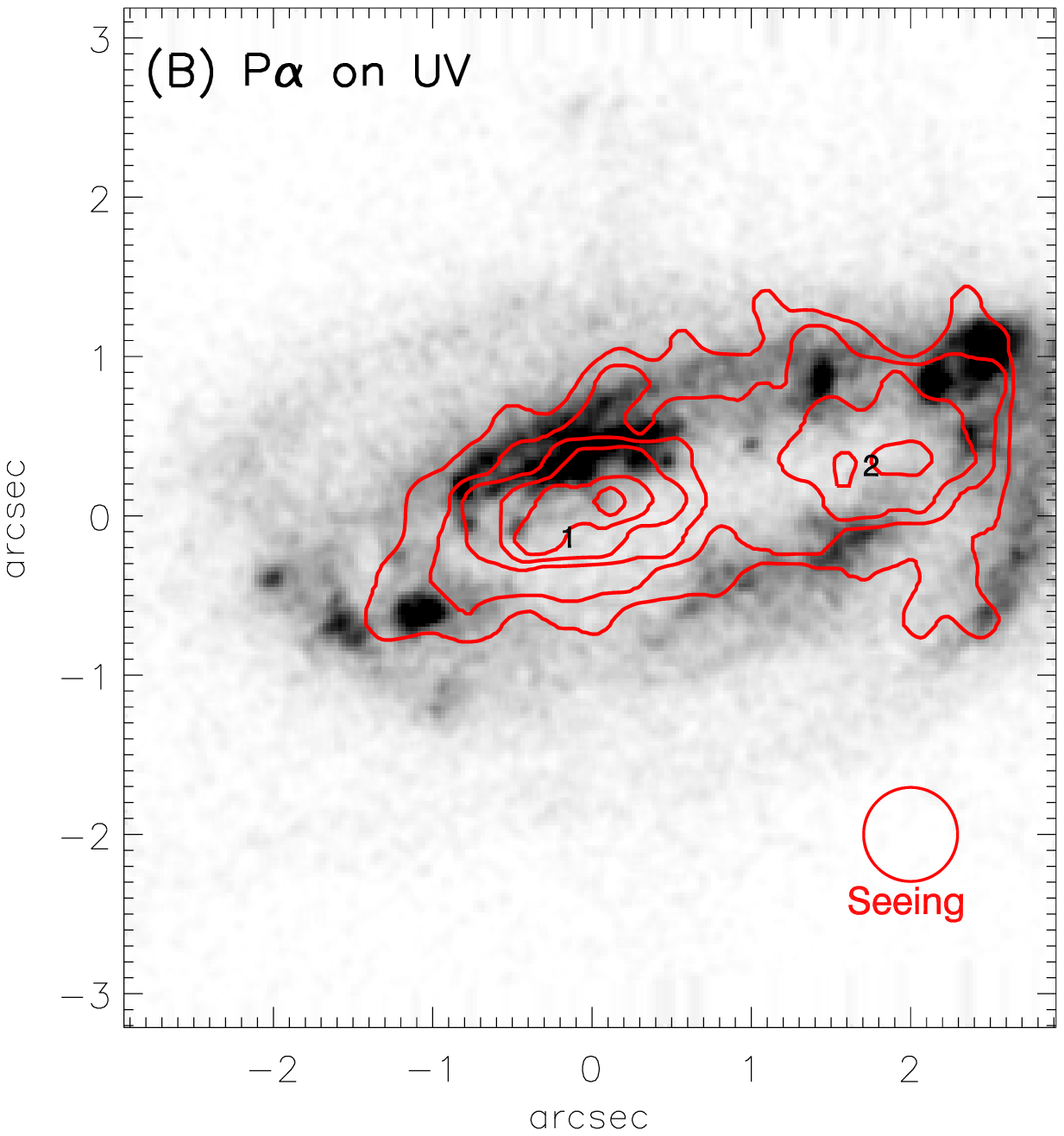,height=8cm}
\caption{Panel (A) presents a HST F814W optical image of Hydra$-$A showing a clear dust 
lane across the centre of the BCG. Contoured over this as solid lines is the S(1) intensity which 
can be seen to run along the length of the dust lane. The dashed lines show the radio emission with 
central point source removed. The radio jets can be seen extending to the north-east and south west of the 
BCG, almost perpendicular to the dust lane in projection. Panel (B) presents a HST F140LP UV 
image of Hydra$-$A which shows a clumpy structure indicating regions of localised star 
formation.  The intensity of the emission from the Pa$\alpha$ line is contoured and shows a 
similar position and structure to the S(1).  A depression in the UV can be seen at 
the position of the dust lane suggesting that some star-formation may be obscured at UV 
wavelengths. The numbers on the two images match the centroids of the two Pa$\alpha$ peaks and the position (0,0) corresponds to the centre of the BCG in optical light.
}
\label{fig:fors}
\end{figure*}

\begin{figure*}
\psfig{figure=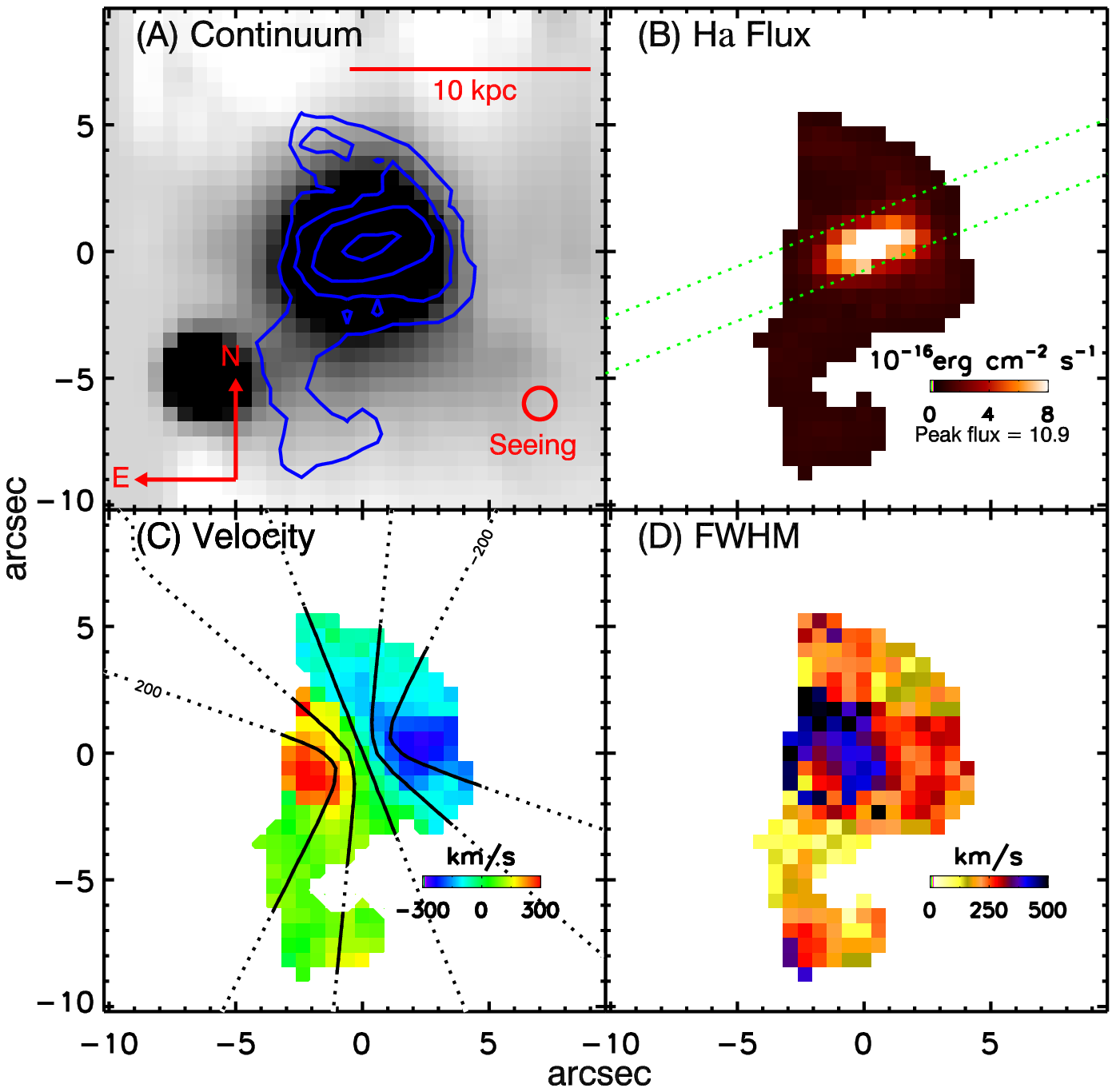,height=18cm}
\caption{
This figure shows the IFU maps of the H$\alpha$ emission as taken from fits to the 
H$\alpha$/[NII] triplet observed in the VIMOS cubes.  Panel (A) shows a 
continuum image made by collapsing the cube, the contours show the H$\alpha$ 
emission clearly centred on the BCG.   
Panel (B) is a H$\alpha$ Flux map which shows a disc of bright emission running 
across the BCG. In panel (C) we show the relative velocity of the 
H$\alpha$ line to the galaxy redshift, a strong velocity gradient of $\sim$ 600 km 
s$^{-1}$ can clearly be seen.  Contoured on this plot are lines 
of constant velocity created by fitting a disc model to the velocity map.  The final panel (D) shows the 
measured Full Width Half Maximum (FWHM) of the line which can be seen to broaden at 
the centre of the velocity gradient.}
\label{fig:ha}
\end{figure*}

In Figure \ref{fig:fors} we show the HST optical (F814W) and UV (F140LP)
images of Hydra-A. The optical image shows the position of the H$_2$(1-0) S(1) 
and radio emission while the UV image has contours of the Pa$\alpha$. It is apparent 
from this image that there is a surface brightness depression running across the 
BCG at $\sim$ -70 degrees to the north-south line. The Pa$\alpha$ line flux is contoured 
on this plot and clearly shows the Pa$\alpha$ emission to be consistent with the 
position and orientation of the dust lane.
The IFU maps for the H$\alpha$ line are displayed in Figure \ref{fig:ha}, we show the 
continuum, line flux, relative velocity and FWHM in panels A,B,C and D respectively.
The VIMOS H$\alpha$ flux map shows a bright elongated region which is coincident with and shares the same 
PA as the Pa$\alpha$ and dust lane as shown in Figure \ref{fig:fors}. Lower surface 
brightness H$\alpha$ emission is also present extended to the north and south of the BCG. 
Although this emission is of lower surface brightness it is still detected at well above 
the 7$\sigma$ and is present in the narrow band image of \citet{mcd10} suggesting it is not 
the result of noise.  Its extent follows the direction of 
the radio jets suggesting the two are possibly related.  In systems with cavities it is 
common to see H$\alpha$ emission associated with the enhanced X-ray emission surrounding cavities 
(for example Abell 2052 in \citet{mcd10}) so this may be a similar effect.

The H$\alpha$ velocity map shows a continuous velocity gradient across the BCG of 
$\sim$ 600 km s$^{-1}$.  This velocity gradient runs across the BCG along 
the major extent of the dust lane.  We 
note that the extended emission to the north and south is slightly blue 
shifted from the velocity of the BCG but is constant and consistent with 
the median velocity of the H$\alpha$ emitting gas.
The line width (all line widths quoted are FWHM deconvolved for instrumental resolution unless 
otherwise stated) of the emission in the central 1.2$\times$1.2 arcsec$^2$
of the BCG is 300 $\pm$ 15 km s$^{-1}$ and falls to 250 $\pm$ 13 km s$^{-1}$ 
at the edges of the region containing the velocity gradient. The lower surface 
brightness extensions to the north and south have lower and more uniform 
line widths of 150 $\pm$ 16 km s$^{-1}$.

\begin{figure*}
\psfig{figure=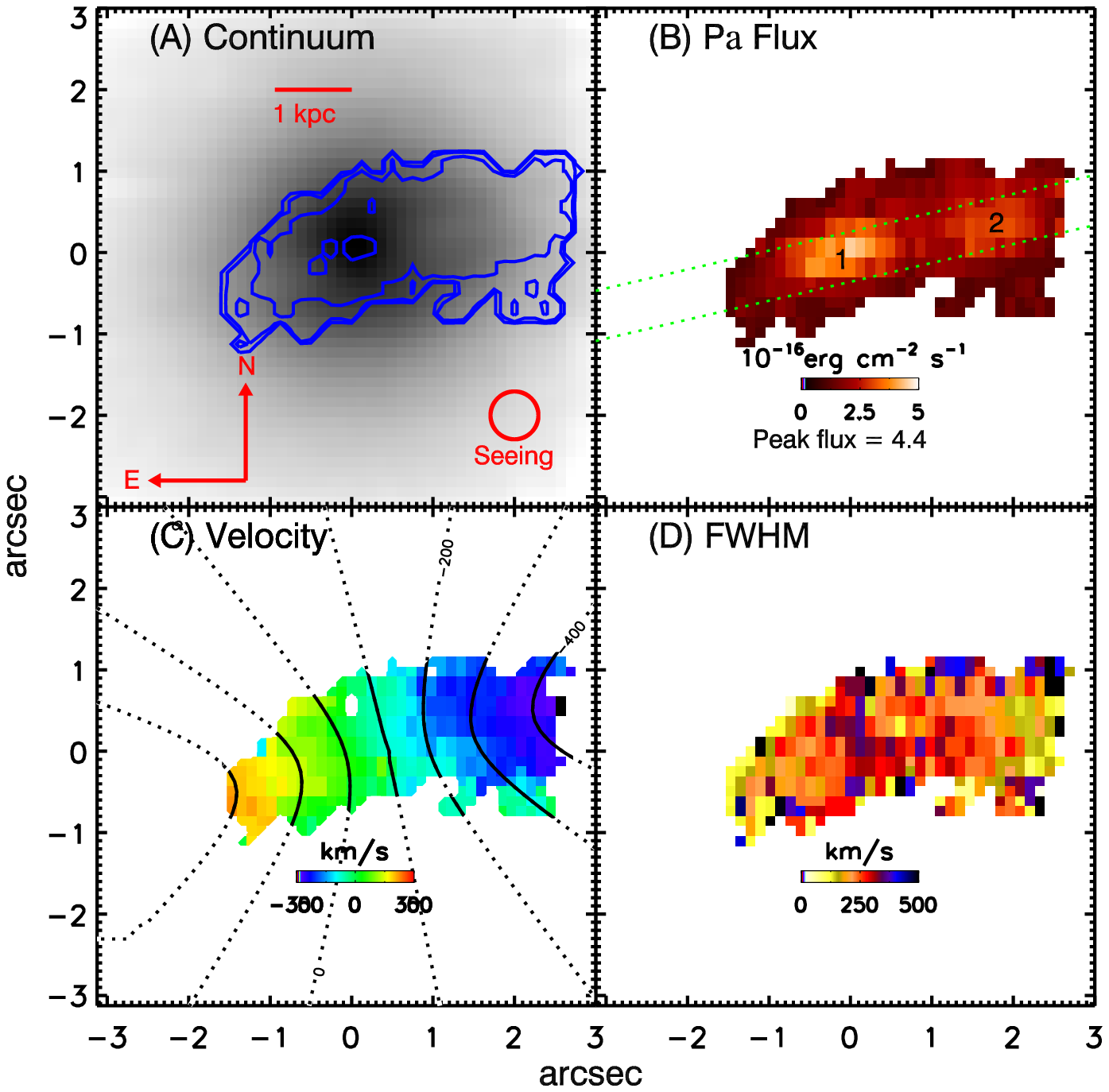,height=18cm}
\caption{
This figure shows the IFU maps of the Pa$\alpha$ emission as taken from fits to the 
Pa$\alpha$ emission line observed in the SINFONI cubes.  Panel (A) shows a 
continuum K-band image created by collapsing the cube with contours of the 
Pa$\alpha$ emission.  
In panel (B) is a Pa$\alpha$ Flux map which shows an elongated structure across the 
BCG. In panel (C) we show the relative velocity of the 
Pa$\alpha$ line to the galaxy redshift; a strong velocity gradient of $\sim 600$ km 
s$^{-1}$ can be seen extending over $\sim$ 4 arcsec (with 1 arcsec $\sim$ 1.05 kpc) 
along a similar PA to the H$\alpha$.  Contoured on this plot are 
lines 
of constant velocity created by fitting a disc model to the velocity map.
  The final panel (D) shows the 
measured Full-Width Half-Max of the line which seems fairly uniform.}

\label{fig:pa}
\end{figure*}

Figure \ref{fig:pa} shows a K-band continuum image produced by collapsing the 
K-band SINFONI observation over a region free of telluric absorption 
(2.07 -- 2.32 $\mu$m).  The dust lane seen in the 
optical image is also present in the NIR but is much less prominent due to the weaker
dust absorption and the emission from a point source at the centre of the BCG.
The Pa$\alpha$ map shows emission elongated across the galaxy in the same 
direction as the dust lane.  The emission shows a two 
component structure with a depression in the Pa$\alpha$ emission separating
two bright components.
Assuming a temperature of the ionised gas of 10$^4$ K the mass of ionised hydrogen in a 
system can be estimated from the Pa$\alpha$ flux
\citep{oon10} as
\begin{equation}
\begin{split}
M_{HII} = & 2.41 \times 10^{18}\left(\frac{F_{Pa\alpha}}{\rm erg \, s^{-1} \, cm^{-2}}\right)\left(\frac{D}{\rm Mpc}\right)^{2} \\
& \times \left(\frac{n_e}{\rm cm^{-3}}\right)^{-1} M\odot
\end{split}
\label{eq:pa}
\end{equation}
where $F_{Pa\alpha}$ is the Pa$\alpha$ flux, $D$ is the distance in Mpc and $n_e$ is the 
electron number density.  Assuming an electron density of $n_e$ = 200 cm$^{-3}$ as in 
\citet{oon10}, a distance of 242 Mpc and the measured Pa$\alpha$ flux of 6.31 $\pm$ 0.39 $\times$ 
10$^{-14}$ erg s$^{-1}$ cm$^{-2}$ we find a HII mass of 4.5 $\pm$ 0.28 $\times$ 10$^7$ M$_{\odot}$.
The Pa$\alpha$ shows a similar velocity structure to the H$\alpha$ with 
a continuous gradient of $\sim$ 500-600 km s$^{-1}$ running across the 
emission. 
The line width (FWHM) of the Pa$\alpha$ is higher in the central 1.2$\times$1.2 
arcsec$^2$ of the emission (330 $\pm$ 39 km s$^{-1}$) than on the edges (150 
$\pm$ 17 km s$^{-1}$).  This is similar to the linewidth variation seen in the 
H$\alpha$ except that the FWHM at the edges is substantially lower. We note however 
that the lower spatial resolution of the VIMOS observations makes the H$\alpha$ FWHM 
more responsive to broadening by the high velocity gradient in this system.
The velocity width seen in the two ionised lines is consistent with observing through a 
rotating disc 
where we sample less random velocity components through the edge of the disc than 
through the centre. However, our current observations are not sufficiently
detailed to fully constrain the emissivity within the disc to determine
if it is continuous or there are gaps in the density that imply that the
gas is in a ring.

\begin{figure*}
\psfig{figure=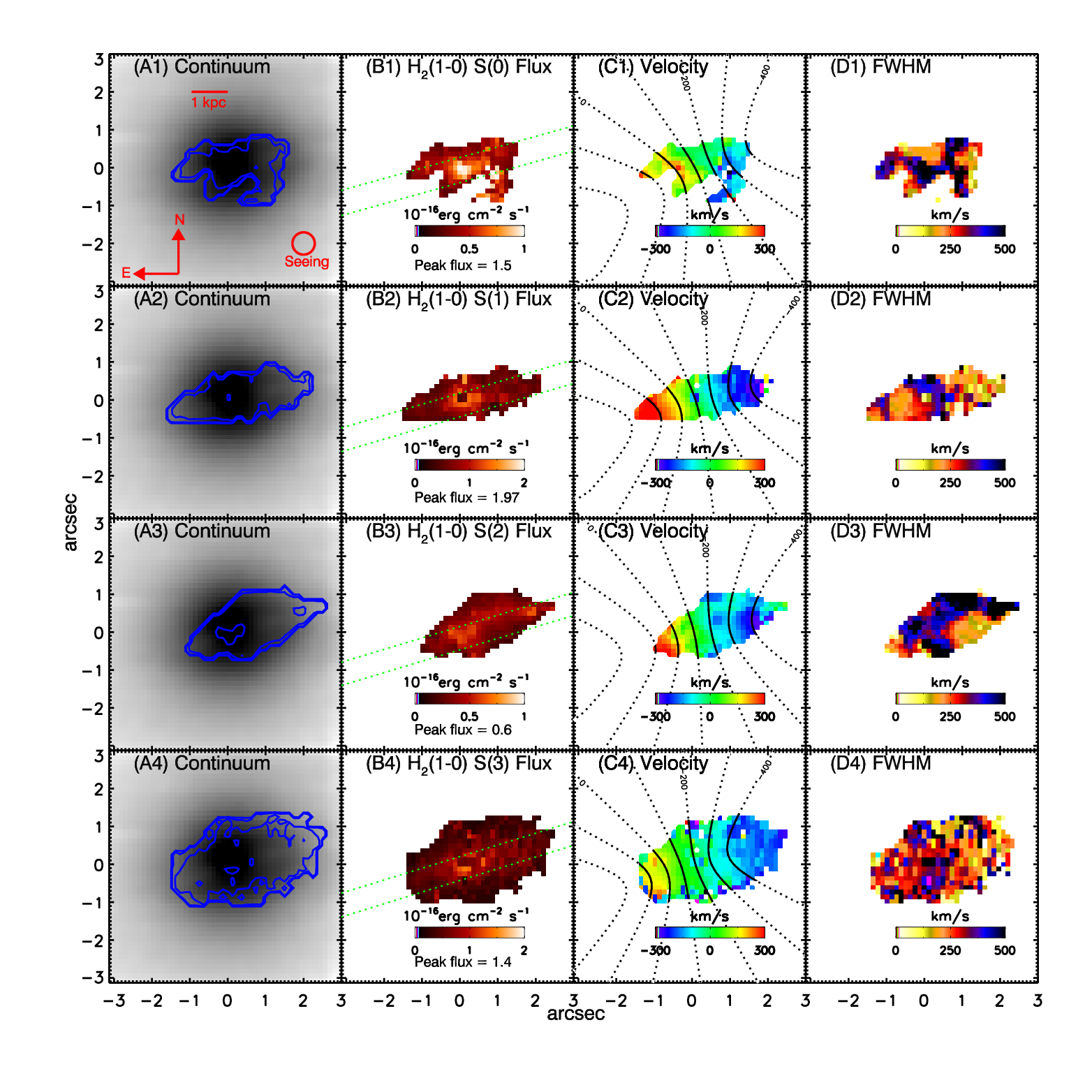,height=18cm}
\caption{
This figure shows the IFU maps of the H$_2$(1-0) S(0) to S(3) emission as taken from fits 
to the emission lines observed in the SINFONI cubes.  From top to bottom the panels are 
arranged in order from S(0) to S(3). The far left panels (A) show a 
continuum K-band image created by collapsing the cube with a flux map for each line 
contoured in blue.  The centre left panels (B) show the flux maps which typically show an 
elongated structure across the BCG.  The dotted lines on these panels show the region 
from which the rotation curves were extracted and represent the position angle 
of the disc as determined by the disc fitting. In the centre right panels (C) we show the 
relative velocity of each emission line to the galaxy redshift. A strong velocity 
gradient of $\sim 600$~km~s$^{-1}$ can be seen along a similar PA to the H$\alpha$.  
Contoured on these plots are lines of constant velocity created by fitting a disc model 
to the velocity maps. The far right panels (D) show the 
measured Full-Width Half-Max of the lines. Each line shows an increase in the FWHM 
at the point where the flux map appears brightest, for the S(1) and S(3) lines this 
point also matches the dynamical centre (the point at which the velocity of the disc 
model goes through 0). However the S(0) and S(2) lines 
also show enhanced line width along the edges of the emission, though this is likely 
a result of the lower signal to noise and therefore not real. The consistency of the 
velocity structure of these lines with those of the ionised H$\alpha$ and Pa$\alpha$ 
suggests that the warm molecular gas occupies the same rotating disc as the ionised gas.}

\label{fig:s0}
\end{figure*}

The H$_2$ (1-0) S series lines sample the warm (1--2 $\times$ 10$^3$ K) 
vibrationally excited molecular gas within the BCG.  Within the 
K-band spectrum of Hydra-A we find four strong lines associated with the 
molecular gas, H$_2$ (1-0) S(0) to S(3) (Figure \ref{fig:s0} shows the fits to all 
four H$_2$ (1-0) S lines).
The S(1) line emission is shown on the second row of Figure \ref{fig:s0}.  It is the most 
reliably characterised of the
molecular lines due to being strong and isolated from other lines and telluric features. 
For 
consistency with the other molecular lines however we present a map of the fits 
to emission above the 5$\sigma$ significance level. The S(1) emission shows a clear 
extent, 
matching that seen in the Pa$\alpha$ and dust lane. In contrast however the 
S(1) flux map does not show the double component structure seen in the Pa$\alpha$.
However, the velocity structure of the S(1) matches that seen in the ionised lines with 
a clearly visible velocity gradient of $\sim$ 600 km s$^{-1}$ running along the 
long axis of the emission.
The FWHM of the S(1) emission peaks at 346 $\pm$ 96 km s$^{-1}$ in the central 
1.2$\times$1.2 
arcsec$^2$ of the BCG. This width is higher than that seen in the Pa$\alpha$ and H$\alpha$
but is consistent within the errors. The width falls to 200 $\pm$ 55 km s$^{-1}$ 
at the edges which is consistent with both the ionised lines.  

The mass of warm molecular hydrogen can be estimated from the H$_2$ (1-0) S(1) flux
using equation 4 from \citet{oon10}, assuming a single kinetic temperature of 
T$_{vib}$ = 2000 K and a transition probability of 3.47 $\times$ 10$^{-7}$ s$^{-1}$, as
\begin{equation}
M_{H_{2}} = 5.08 \times 10^{13} \left(\frac{F_{H_{2}(1-0)\ S(1)}}{\rm erg \, s^{-1} \, cm^{-2}}\right)\left(\frac{D}{\rm Mpc}\right)^{2} M\odot
\label{eq:s1}
\end{equation}
where $F_{H_{2}(1-0)\ S(1)}$ is the flux in the S(1) line, and $D$ is the distance in Mpc.
Again using a distance of 242 Mpc and the measured $F_{H_{2}~(1-0)~S(1)}$ flux of 142 $\pm$ 27 $\times$ 
10$^{-16}$ erg s$^{-1}$ cm$^{-2}$ we find a H$_{2}$ mass of 4.2 $\pm$ 0.8 $\times$ 10$^4$ M$_{\odot}$.

The S(3) line (Figure \ref{fig:s0}, row four) is another strong molecular line. However, its 
position in the spectrum places it near a strong telluric feature.  The removal of 
this telluric feature results in an increase in noise compared to other parts of the 
spectrum.  While the line is clearly detected at greater than the 5$\sigma$ 
significance limit, the physical properties derived from the fit are less reliably 
determined than 
those of the S(1) line.  We also note that 
in the nuclear regions the S(3) emission appears to be blended with HeI emission,  
requiring that the two lines have been fitted together with a fixed rest wavelength 
separation.
The velocity profile of the S(3) line is consistent with that from the other molecular 
lines 
and ionised lines presented so far, with a velocity gradient of $\sim$ 500--600 
km s$^{-1}$ running along the direction of the dust lane.
The linewidth of S(3) is mostly consistent with that of S(1) peaking at 375 $\pm$ 
80 km s$^{-1}$ in the central 1.2$\times$1.2 arcsec$^2$ of the emission and falling to 
230 $\pm$ 72 km s$^{-1}$ at the edges.

Of the molecular hydrogen lines, S(0) is the least reliable since it is located at 
the noisy red edge of the spectrum.  As such, fits could only be obtained at 
the 5$\sigma$ significance level for this line.

The S(0) fits (Figure \ref{fig:s0}, row one) do not show as extreme a velocity gradient 
as seen in the 
Pa$\alpha$ but they do show a similar velocity structure with a continuous 
velocity gradient running almost east to west across the emission.  We 
note that the S(0) emission is not as extended (likely due to lower signal--to--noise) 
along this direction as the other K-band lines which may account for this.  
The FWHM of the S(0) line is highest at a position centred on the peak of the 
emission, which is consistent with being at the centre of the BCG. The FWHM in 
the central 1.2$\times$1.2 arcsec$^2$ is 308 $\pm$ 104 km s$^{-1}$. However, 
the map shows regions of similar linewidth throughout the structure.  The similarity
of linewidths is possibly due to blending with noise spikes that artificially broaden 
the line fit. 

The S(2) emission is shown in row three of Figure \ref{fig:s0}.  While the S(2) emission 
from Hydra-A was weaker than the S(0) its observed wavelength put it in a 
significantly less noisy part of the spectrum resulting in the detection of more 
emission at 5$\sigma$.  As such 
its structure more closely matches that of the S(1), suggesting the structure in 
the S(0) map is likely the result of incompletely detected emission  
 and is not real.
As expected from the other molecular lines the velocity structure of the S(2) 
emission matches that of the ionised gas with a $\sim$ 600 km s$^{-1}$ gradient 
running along the line of the dust lane.
The linewidth of S(2) appears considerably higher than the S(1).  This is however 
likely a result of the binning used by the fitting procedure to obtain the required 
signal to noise. By binning together pixels at substantially different velocities 
before fitting the line width can be artificially increased.  The velocity gradient in 
the central regions is $\sim$ 30 km s$^{-1}$ per pixel, thus binning 3$\times$3 pixels 
can result in a broadening of the line on the order of 100 km s$^{-1}$ which is 
consistent with the difference in line width observed. Consulting the 
integrated spectrum of the 1.2$\times$1.2 arcsec$^2$ centred on the BCG suggests that the 
linewidth is consistent with that of the other lines (362 $\pm$ 70 km s$^{-1}$).
As such there is no significant difference in the 
warm molecular line widths as is expected since they are emitted from the same 
gas. 

The separate component seen in the Pa$\alpha$ map does not appear in the 
(1-0) S series line emission.  This non-detection is likely to be due to the lower 
signal--to-noise 
as the lines are present within the total spectrum of this region (extracted 1$\times$1 
arcsec$^2$ centred on the offset Pa$\alpha$ peak)
though the line is weak compared to Pa$\alpha$.

\begin{figure*}
\psfig{figure=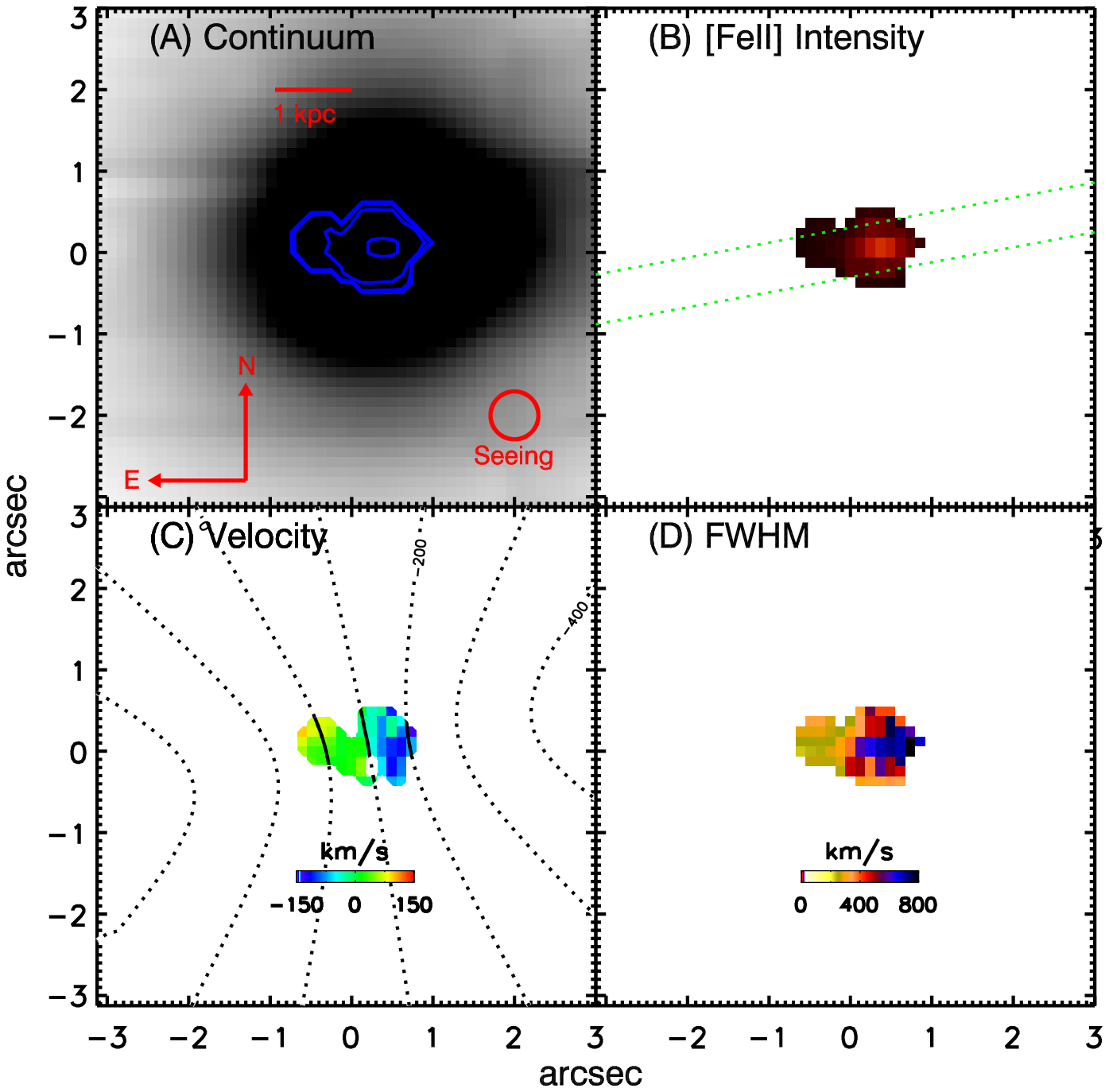,height=18cm}
\caption{
This figure shows the IFU maps of the [FeII] emission as taken from fits to the 
[FeII] emission line observed in the SINFONI cubes.  Panel (A) shows a 
continuum K-band image created by collapsing the cube with contours of the 
[FeII] emission.  
In panel (B) is a [FeII] intensity map, it can be seen that the [FeII] emission is 
much more centrally 
concentrated than the other lines, which is expected as it is associated
with the highest density regions.   
In panel (C) we show the relative velocity of the 
[FeII] line to the galaxy redshift, a strong velocity gradient of $\sim 200$ km 
s$^{-1}$ can be seen along a similar PA to the H$\alpha$.  Contoured on this plot are 
lines 
of constant velocity created by fitting a disc model to the velocity map.  It should be 
noted that the [FeII] emission is barely resolved at just twice the seeing so this 
velocity gradient may not be real.
  The final panel (D) shows the 
measured Full-Width Half-Max of the line which shows a much broader profile in the 
central regions than the other lines.}

\label{fig:fe}
\end{figure*}

[FeII] emission was the only line detected in the H-band observations. The maps 
presented in Figure \ref{fig:fe} show that the [FeII] emission is compact and 
located at the centre of the BCG. The luminosity of [FeII] emission has a high 
dependence on the gas density \citep{bau94} so we would expect it to be brightest 
in the central regions where the gas density is higher.
Despite being compact the line does appear to be extended to the east on 
scales slightly greater than the seeing.  
Within this small extent there appears to be a velocity change of 
$\sim$ 200 km s$^{-1}$ across the emission.  This velocity gradient is 
consistent with that found from the other ionised and molecular lines 
across a similar extent.  However, higher resolution observations of this
emission are required to confirm if this velocity gradient is real.  If it is, 
it presents a direct link between the gas on galaxy--wide scales 
and the nuclear region. 
The peak FWHM of the [FeII] line is greater than that seen in the other lines, 
at $\sim$ 700 km s$^{-1}$, as would be expected if this emission came from the 
denser nuclear region.  However, we note that the error on the line width is 349 km 
s$^{-1}$ in this region.  By contrast the component extended to the east has a 
substantially lower width ($\sim$ 300 $\pm$ 55 km s$^{-1}$), which is more consistent 
with that seen in the molecular emission lines.


\begin{table*}
\begin{center}
\scriptsize
\centerline{\sc Table \ref{tab:info}.}
\centerline{\sc Properties of the seven emission lines fit with disc models}
\smallskip
\begin{tabular}{| c ||  c |  c |  c |  c |  c | c |}
\hline
\noalign{\smallskip}
Line & Position angle & Inclination & Transition R & Luminosity & V$_{2.2}$ & Redshift\\ 
 & (Degrees) & (Degrees) & (kpc) & ($10^{40}$ erg s$^{-1}$) & (km s$^{-1}$) & (At zero velocity) \\ \hline 

H$\alpha$ & -74 $\pm$ 3.0 & 56 $\pm$ 2.3 & 1.68 $\pm$ 0.26 & 14.22 $\pm$ 0.76 & 397 $\pm$ 49 & 0.05439[12] \\ 
Pa$\alpha$  & -77 $\pm$ 3.3 & 69 $\pm$ 2.0 & 1.97 $\pm$ 0.23 & 45.24 $\pm$ 2.83 & 341 $\pm$ 36 & 0.05452[8] \\ 
H(1-0) S(0) & -66 $\pm$ 2.9 & 65 $\pm$ 2.0 & 1.46 $\pm$ 0.20 & 5.16 $\pm$ 1.11 & 384 $\pm$ 25 & 0.05439[5] \\ 
H(1-0) S(1) & -74 $\pm$ 2.6 & 64 $\pm$ 2.9 & 1.48 $\pm$ 0.27 & 15.58 $\pm$ 5.94 & 381 $\pm$ 65 &  0.05440[5] \\ 
H(1-0) S(2) & -78 $\pm$ 5.3 & 64 $\pm$ 3.6 & 1.76 $\pm$ 0.21 & 7.69 $\pm$ 1.14 & 369 $\pm$ 42 & 0.05438[6] \\ 
H(1-0) S(3) & -73 $\pm$ 3.0 & 66 $\pm$ 2.7 & 1.54 $\pm$ 0.21 & 25.80 $\pm$ 1.34 & 421 $\pm$ 66 & 0.05439[3] \\ 
{[}FeII] & -80 $\pm$ 0.9 & 56 $\pm$ 1.8 & 1.70 $\pm$ 0.20 & --- & 355 $\pm$ 33 & 0.05473[10] \\ 

\end{tabular}
\caption{[NII] and H$\alpha$ have the same PA,inclination and velocity as they are fit as a 
triplet. The transition radius is measured as the point at which the fit rotation curve begins to flatten.  A suitable standard star for the [FeII] was not taken so no value of the luminosity could be obtained. V$_{2.2}$ is defined as the inclination corrected 
velocity at R$_{2.2}$, with 
the values of R$_{2.2}$ set to 3.2 and 2.20~kpc for VIMOS and SINFONI, respectively. 
The square brackets give the error on the last decimal place.}
\label{tab:info}
\end{center}
\end{table*}

\subsection{Rotation of the ionised gas}

\begin{figure}
\psfig{figure=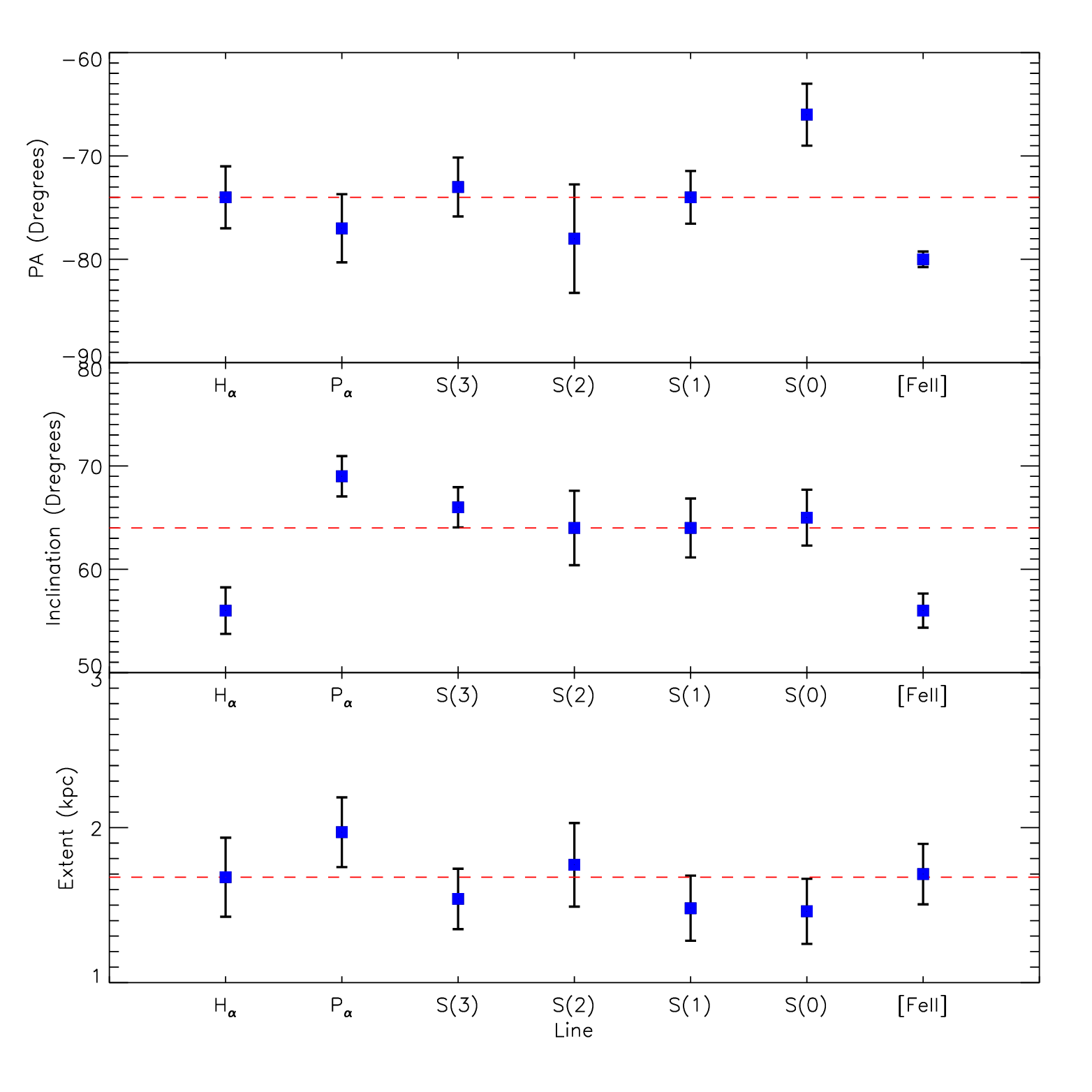,height=8cm}
\caption{Comparison of the key parameters of the best fit for the disc model for each line (Note that the rotation velocity comparison is shown later in Figure \ref{fig:vrot}).  The top panel shows a comparison of the position angle, the middle shows the inclination angle and the bottom shows the distance at which the velocity curve turns over.  The dashed line on each plot shows the median value.  The errors are calculated from the Monte Carlo fitting routine and are the maximum deviation from the best fit parameters with a $\Delta\chi^2<1$ one assuming all other parameters are allowed to minimise. Note the smaller error on the [FeII] is a result of having to use more constrained limits on the parameters in order to get a fit.  For each of these three parameters the disc models show a good consistency between the lines.}
\label{fig:parcomp}
\end{figure}

The ionised and warm molecular gas emission lines all show kinematics which are 
consistent with an ordered rotation. 
Disc models were fit to the velocity maps shown above in order to determine the 
consistency of this rotation between the different lines.  These disc models 
were constructed such that the velocity profile follows an arctan function. 
While this model is not physically motivated, it was found to be a good approximation of the velocity 
curve of disc galaxies by \citet{cou97}.
Table \ref{tab:info} gives the 
position angle, inclination and transition radius of the disc as measured by these fits.
The position angle is measured from the north--south line counterclockwise to the blue 
shifted end of the disc.  The inclination is the angle of the plane of the disc relative 
to the plane of the sky with 0$^\circ$ corresponding to a face on disc and 90$^\circ$ 
being an edge on disc.  Finally the transition radius is the point at which the rotation 
curve begins to flatten.
The fits find position angles 
consistent to within $\sim 15^\circ$ (Figure \ref{fig:parcomp}) and rotation velocities consistent to within 
$\sim$60 km~s$^{-1}$ (within the spectral resolution of both instruments). The centre of rotation of each
disc model differs slightly but all are consistent within the seeing of the observations ($\sim$ 0.5 arcsec). 
Given the small spatial scale of the [FeII] emission compared to the
seeing, there are very few uncorrelated data points in the disc fit so
we constrained the disc model for [FeII] to have an inclination and
PA consistent with the other lines. 
We use the stellar sodium--D absorption feature ($\lambda_{rest}$ 5895.92, 5889.95 \AA) from a 
FORS observation to determine the central redshift of the BCG.  From this fit we 
find the stellar component has a redshift of $z=$0.05434[38], which is consistent 
within errors with the redshift at zero velocity for all lines.
Combining the 
information from the seven lines suggests the presence of a disc of ionised and 
molecular gas in the centre of the Hydra-A BCG which is rotating with a 
peak-to-peak velocity 
of $\sim$ 700 km s$^{-1}$ separated by 5 $\pm$ 0.7 kpc. 

\begin{figure*}
\begin{minipage}{0.33\linewidth}
\centering
\psfig{figure=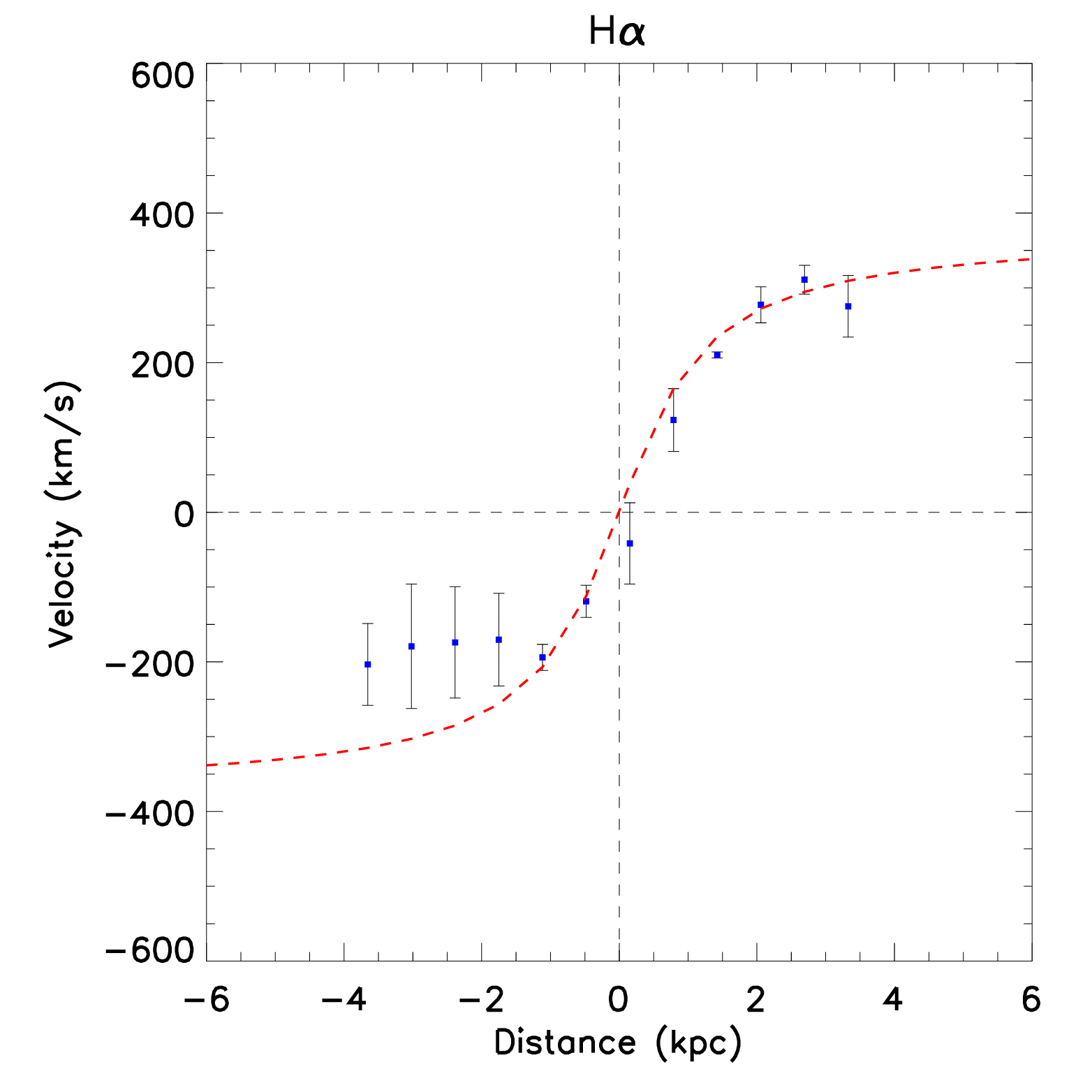,height=6cm}
\psfig{figure=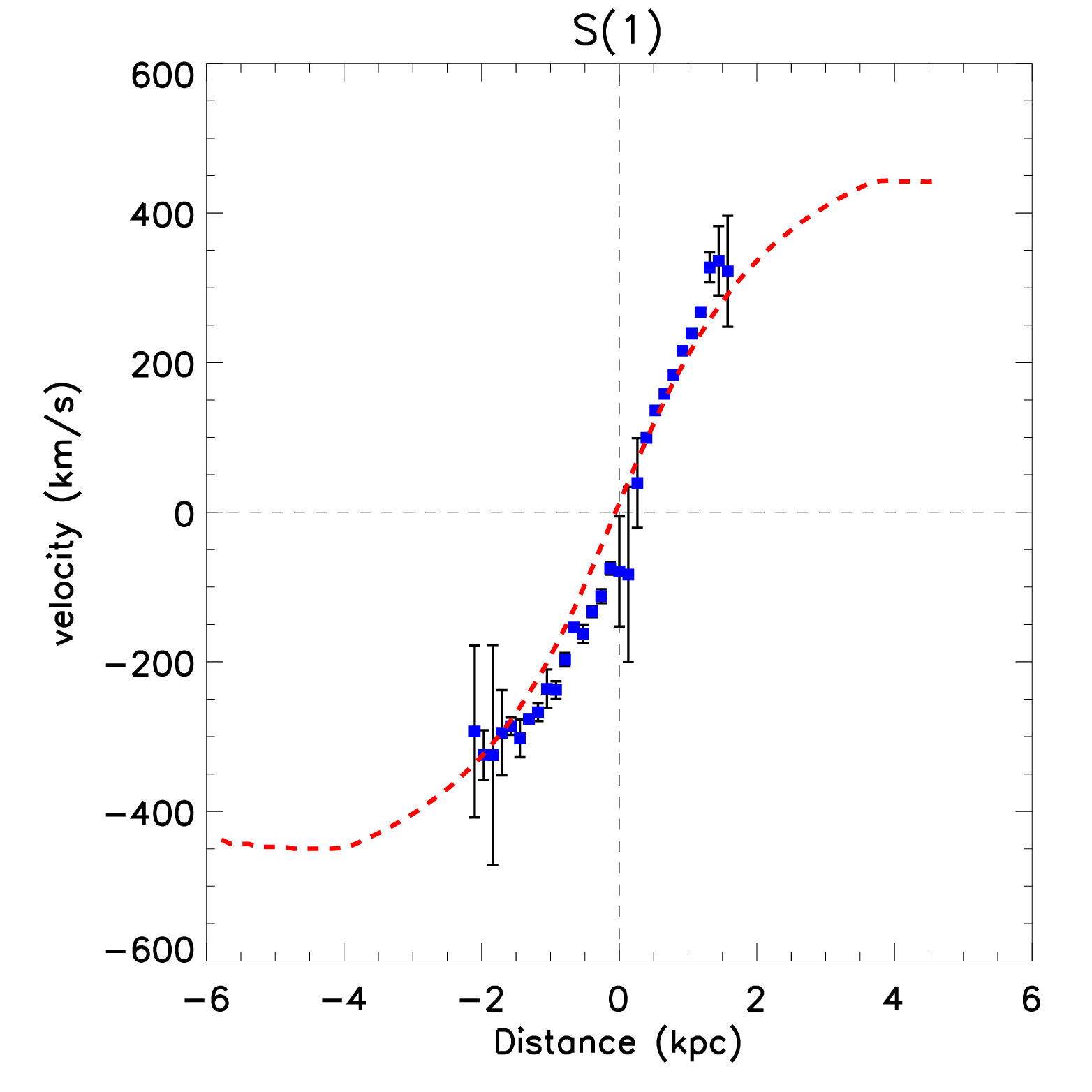,height=6cm}
\end{minipage}
\begin{minipage}{0.33\linewidth}
\centering
\psfig{figure=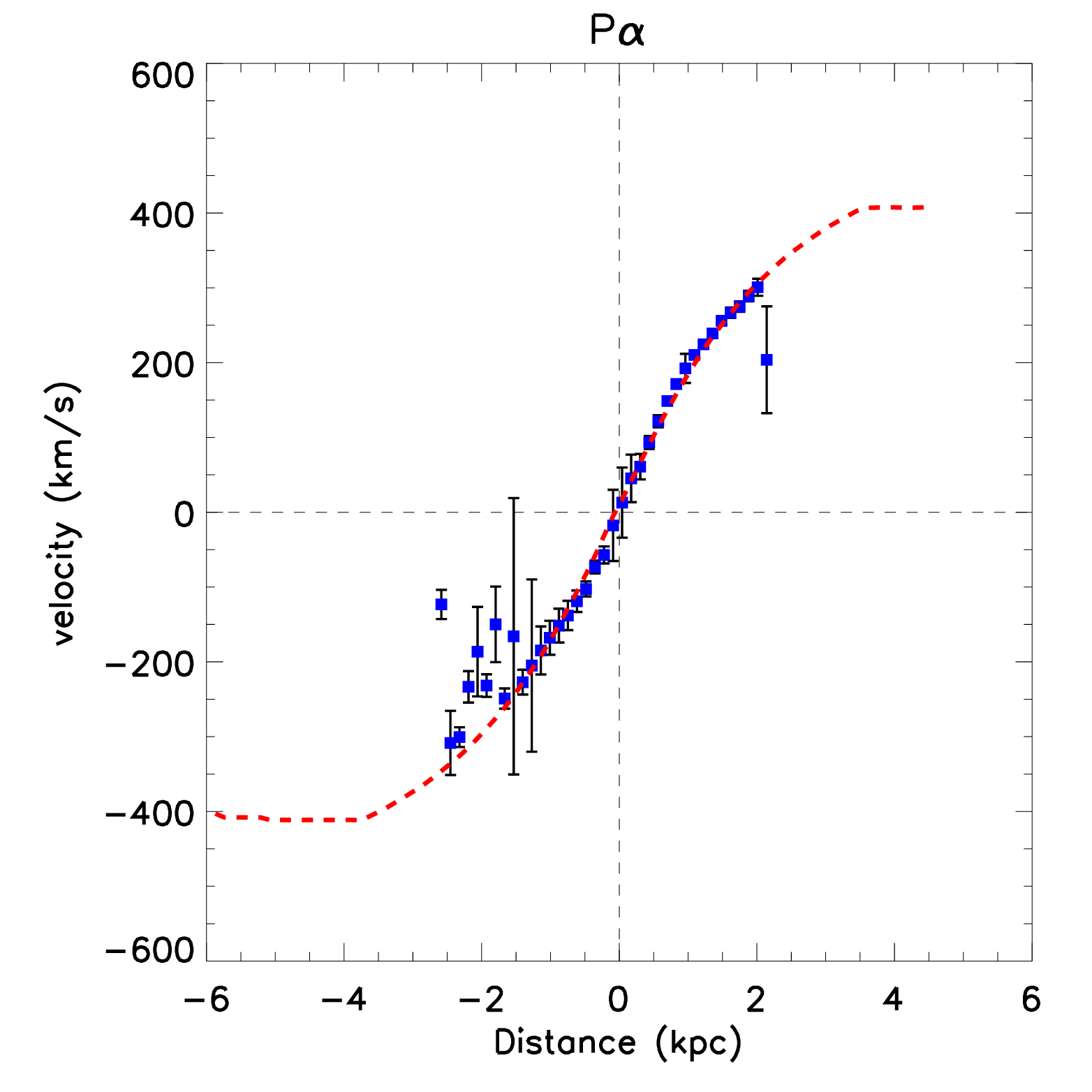,height=6cm}
\psfig{figure=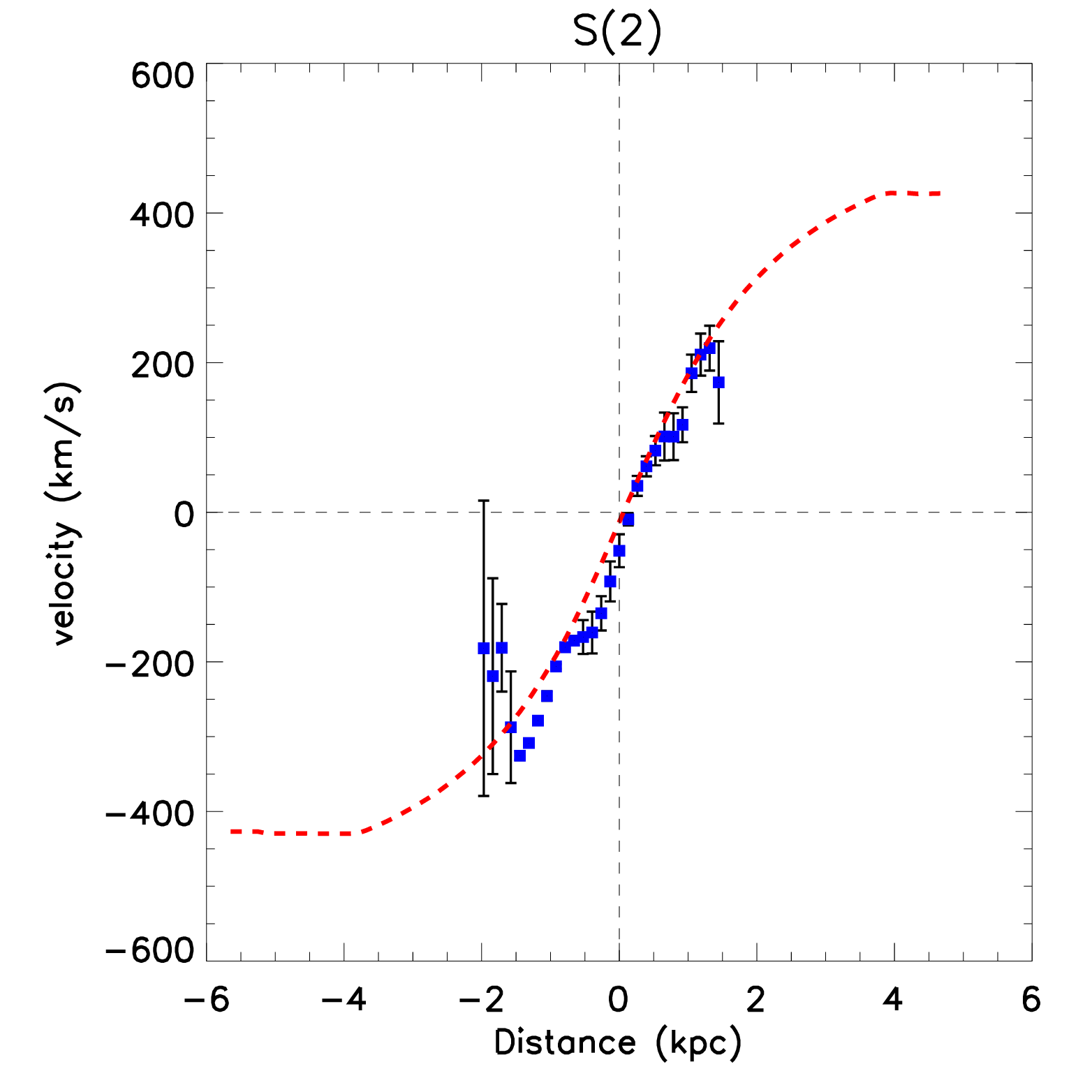,height=6cm}
\end{minipage}
\begin{minipage}{0.33\linewidth}
\centering
\psfig{figure=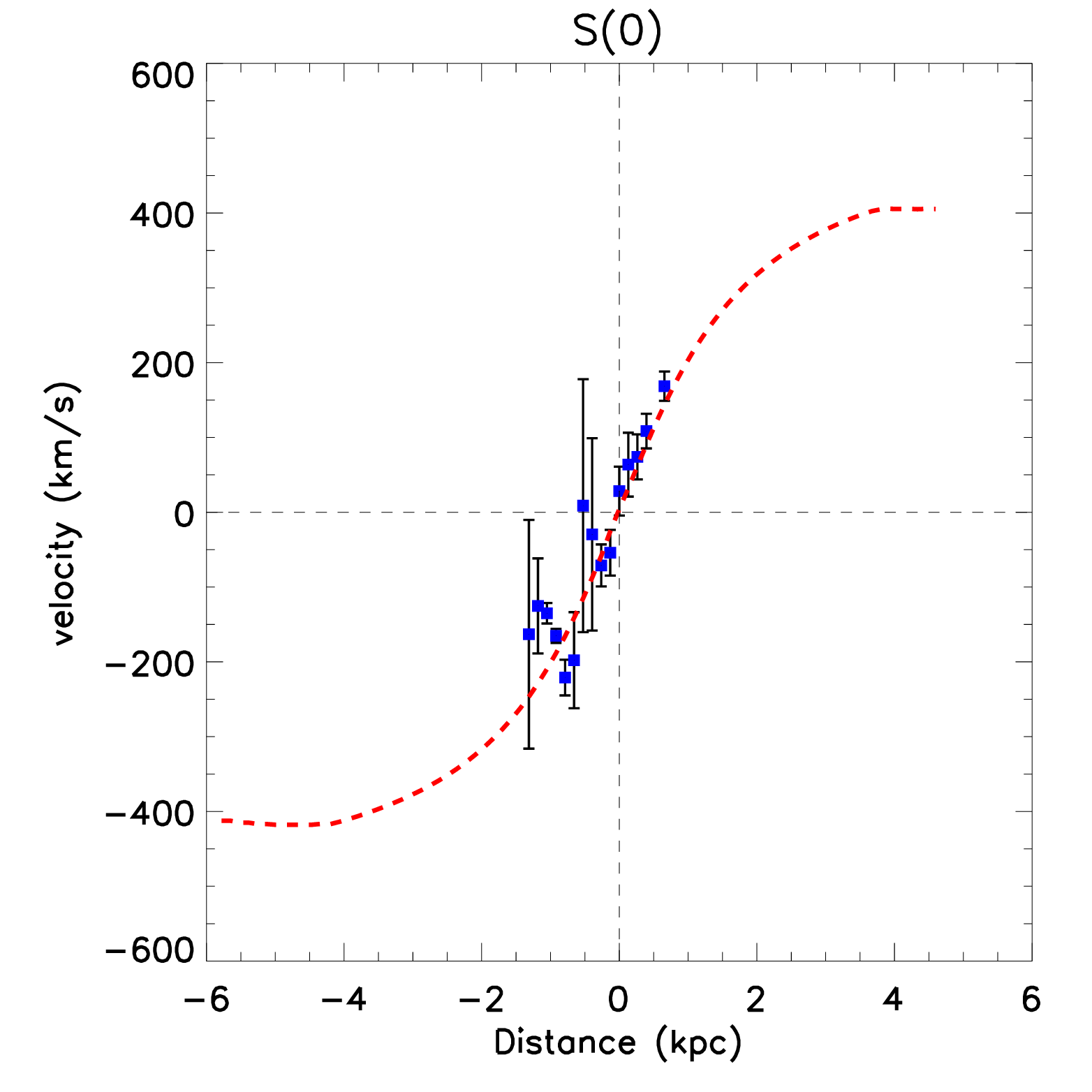,height=6cm}
\psfig{figure=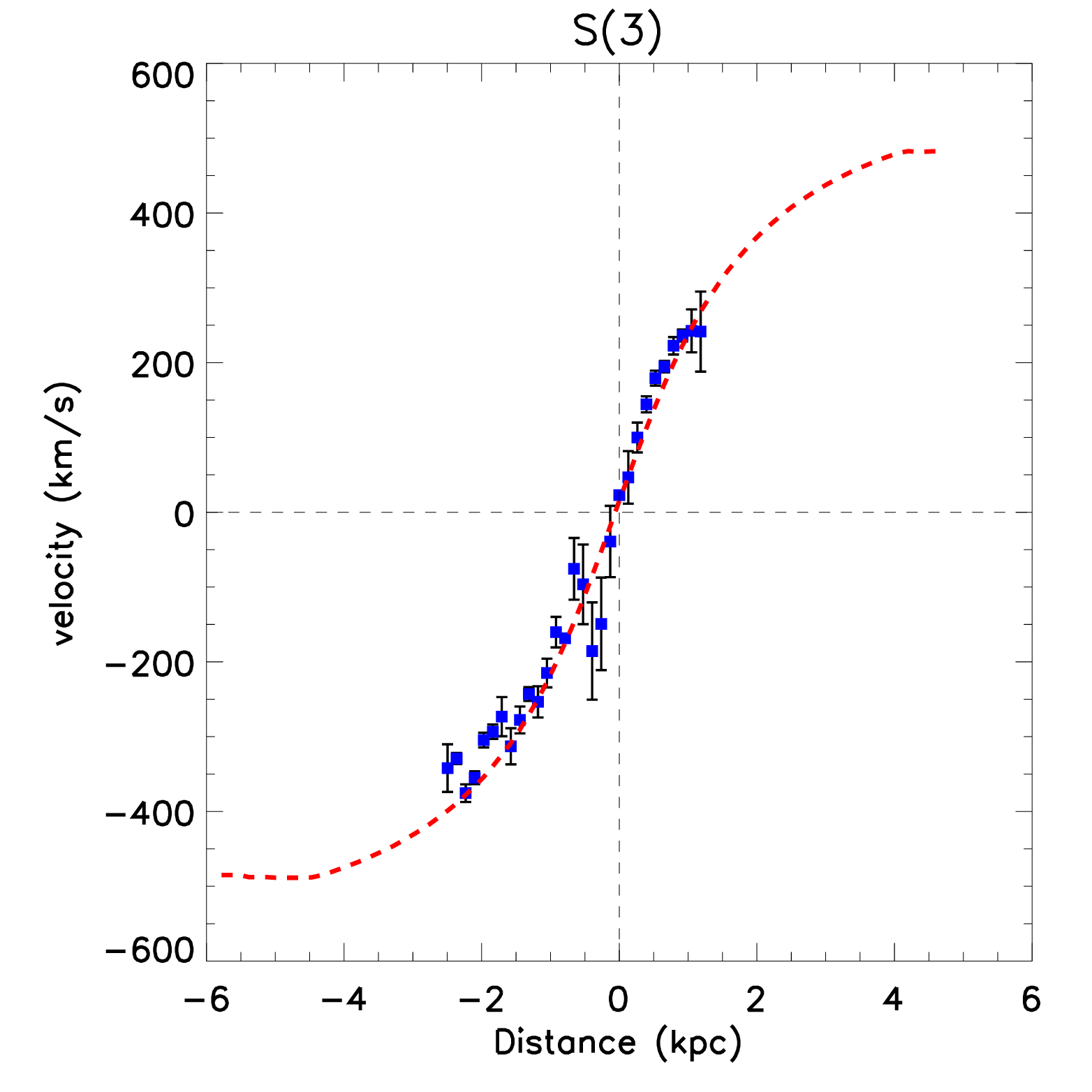,height=6cm}
\end{minipage}
\caption{
Velocity profiles for the six brightest lines from our VIMOS and SINFONI observations.  The profiles are 
presented as observed (not inclination corrected), the errors are 1$\sigma$. The errors on the position (x-axis) are dominated by the seeing, these are not shown but are $\sim$ 1.3 kpc for VIMOS (H$\alpha$) line and $\sim$ 0.5 kpc for SINFONI (other lines).  The dashed 
line shows the velocity profile of the disc model taken along the same line as the data.  
For each line the model provides a good fit to the data suggesting the velocity is well 
fit by a disc model, however we note that for none of the lines do we reach the flattened 
part of the velocity profile.}
\label{fig:rotprof}
\end{figure*}

Figure \ref{fig:rotprof} shows rotation curves extracted from the velocity maps 
along with profiles derived from the models using 
the same PA estimate, which was consistent for the disc models for each line.
The extracted rotation curves all show little deviation from the models except at 
the edges where the line fits are not as precise due to the lower signal--to--noise. 
The S(0) line velocity profile appears less consistent with the model than the 
velocity profiles of the other molecular lines displayed in Figure \ref{fig:rotprof} 
and samples less of the velocity curve. However, 
as discussed in Section \ref{sec:maps} the map of this line is noisier than the others
and it is consistent with the model within errors.  The H$\alpha$ profile appears to be 
slightly offset from the model in the central regions ($\sim$ 0.5 kpc at 0 km s$^{-1}$).
However, we note that this is well within the positional error introduced by the poorer 
seeing of the VIMOS observation ($\sim$ 1.3 kpc).  Due to the [FeII] line being barely 
resolved we do not show its rotation curve.  We note that the rotation curves in the 
inner $\sim$ 2 arcsec (2.1 kpc) appear rigid.  Some caution is required in interpreting 
this as the seeing is on the order of 0.5 arcsec so this could be the effect of a smoothing
of the rotation curve over these scales. However, it is suggestive that the mass density 
in the inner 2 kpc may be nearly constant rather than being proportional to the distance 
as it appears to be in the outer regions (beyond 2 kpc).

\begin{figure}
\psfig{figure=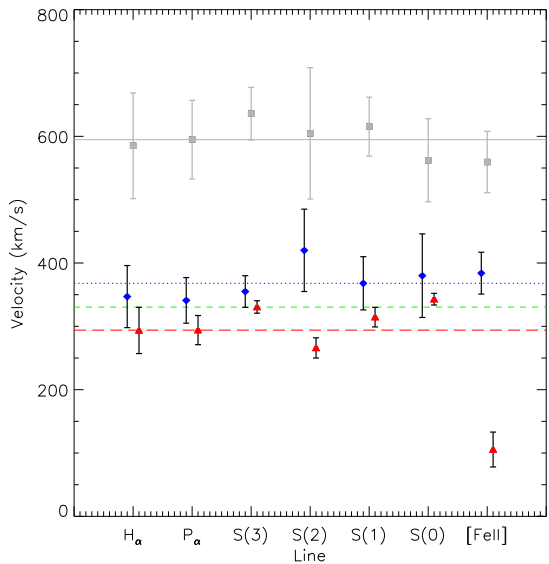,height=8cm}
\caption[Comparison of the rotational velocity predicted by the disc fits for each line.]{
Here we present a comparison of the rotation velocity predicted by the disc fits 
for each line.  The red triangles show the magnitude of the maximum velocity 
measured from the data, the long dashed red line shows the median of these values 
while the dashed green line shows the median corrected for the median inclination.
The blue diamonds show the values of V$_{2.2}$ derived from the models with the dotted 
blue line set at the median V$_{2.2}$.
The squares show the asymptotic velocity from the best fit model for each line with 
the solid line representing their median.  All errors bars are to 1$\sigma$ and it can 
be seen that within errors the value of V$_{2.2}$ are consistent with the inclination 
corrected maximum measured velocity.
}

\label{fig:vrot}
\end{figure}

We use these rotation curves to directly constrain the 
dynamical mass of the BCG within the radius enclosing the disc.  
Assuming the disc is purely exponential then its surface brightness at radius r 
$\Sigma_d$(r) is given by
\begin{equation}
\Sigma_d(r) = \Sigma_0 e^{-{r} \over {r_s}}
\label{eq:r22}
\end{equation}
where $\Sigma_0$ is the central surface brightness and r$_s$ is the disc 
scale length.   
If the disc were purely exponential, the peak rotational amplitude should 
occur at a distance 2.2 $\times$ the disc scale length (r$_s$) \citep{fre70,bin87}.
Using the inclinations calculated above we produce surface brightness profiles of 
the VIMOS and SINFONI maps and fit lines of the form given in equation \ref{eq:r22}
to determine the disc scale length.
We then define the 
position $R_{2.2}$ as a distance from the centre of rotation equal to 
2.2\,$\times$\,the disc scale length and calculate the mass within this region 
\citep[as in][]{mlr11}.  
The inclination--corrected velocity at this point ($V_{2.2}$) is then used to 
calculate the dynamical mass within the disc.  Figure \ref{fig:vrot} compares 
the calculated value of $V_{2.2}$ for each line with the magnitude of the maximum 
velocity measured from the data. It can be seen that once corrected for inclination 
the measured values agree with the calculated values of $V_{2.2}$ indicating that 
the data samples the disc out to $\sim$ $R_{2.2}$.
For the H$\alpha$ emission we 
find $R_{2.2}$ = 3.2 kpc and $V_{2.2}$ = 397 km s$^{-1}$ which corresponds to 
a dynamical mass of 1.2\,$\times$\,10$^{11}$\,M$_{\odot}$. However, we note that 
the spatial resolution of the H$\alpha$ map is 0.6 kpc.  By contrast the Pa$\alpha$ 
map has a spatial resolution of 0.26 kpc and we find $R_{2.2}$ = 2.2 kpc, $V_{2.2}$ 
= 340 km s$^{-1}$ and a dynamical mass of 6.0\,$\times$\,10$^{10}$\,M$_{\odot}$.
As the Pa$\alpha$ map has a higher spatial resolution we use this as $R_{2.2}$, this 
is consistent with the radius of the HI disc found by \citet{dwa94}. 
We thus find the dynamical mass within $R_{2.2}$ to be
$M_{dyn,2.2}$ = 6.0\,$\times$\,10$^{10}$\,M$_{\odot}$.

\subsection{Cold molecular emission}
\label{sec:co}
\begin{figure}
\psfig{figure=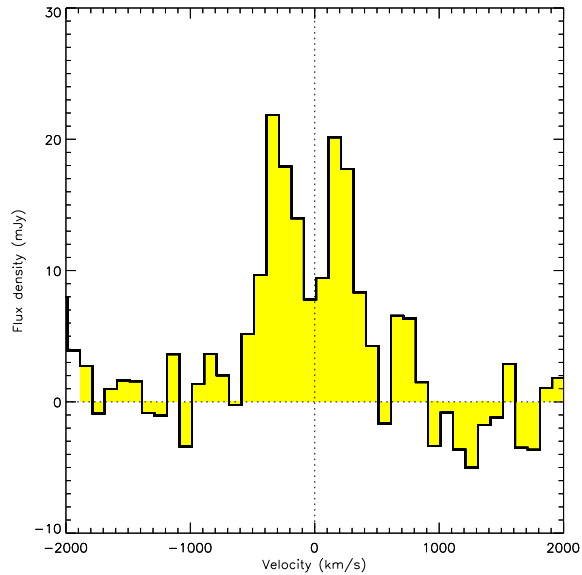,height=8cm}
\caption{CO(2-1) spectrum of Hydra-A taken from the WILMA backend on IRAM 
30m. The line shows a width of 600 to 700 km~s$^{-1}$ consistent with the rotation 
velocity seen in ionised lines. A double peak structure is present with the trough 
between peaks situated at the dynamical centre of the system}
\label{fig:co}
\end{figure}

In Figure \ref{fig:co} we show the CO(2-1) spectrum of Hydra-A which shows a 
strong and broad line present within the cluster.  
This detection is notable for several reasons, not least the 
remarkable width of the line. The full width zero intensity of 1200~km~s$^{-1}$
is comparable to the bandwidth of the past attempts to 
detect CO in this system \citep{ode94,edg01,sc03}
so it would have been impossible to  detect the 
full line flux before the 
advent of the wide bandwidth receivers such as EMIR. The line is 
broader than all previous CO detections in BCGs 
by a factor of 30 per cent
and it suggests there may be a potentially significant number of systems that have
escaped detection due to their intrinsic line width.

The spectrum also points to considerable velocity structure within the line. There
are only three published double-peaked CO detections in BCGs (3C31 and 3C264 in \citet{lim00} 
and 3C84 in \citet{sal11}) so the discovery of a third in a 3C
radio galaxy points to a potential connection between powerful
radio emission and cold gas discs. This connection extends to other
radio galaxies, such as 3C236 \citep{lab12} where a molecular
gas disc is found.

The total CO(2-1) line intensity is 1.87$\pm$0.24 K~km~s$^{-1}$ in units of
measured antenna temperature ($T_A^*$) or 2.79$\pm$ 0.36 K~km~s$^{-1}$ in units of 
main beam temperature using the conversion factor of 1.49 at the observed frequency of 
CO(2-1) (218.545 GHz).  Following \citet{sol97} the CO luminosity is given by

\begin{equation}
L'_{CO} = 23.5 I_{CO} \Omega_s*b {{D_L^2}\over {(1+z)^3}}
\label{eq:col}
\end{equation}

with L$'_{CO}$ in K~km~s$^{-1}$~pc$^2$. Here I$_{CO}$ is the main beam temperature, 
$\Omega_s*b$ is the solid angle of the source convolved with the beam and D$_L$ is the 
luminosity distance of the source.  Assuming that the CO emission comes from the 
region seen in the IFU maps shown in Section \ref{sec:maps} then the source is 
much smaller than the beam and we can use the approximation 
$\Omega_s*b$~$\propto$~$\Omega_b$ which is 144 arcsec$^2$ for CO(2-1).  The source 
redshift is z$=$0.054878 which implies a luminosity distance of D$_L$~$=$~242 Mpc. 
This gives a CO luminosity of 
$L'_{CO}$~$=$~$4.71 \pm 0.61 \times 10^{8}$~K~km~s$^{-1}$~pc$^2$ and assuming the 
CO is thermally excited then $L'_{CO(1-0)} = L'_{CO(2-1)}$ (which is consistent with our 
marginal detection of CO(1-0)).  We then compute the 
H$_2$ mass using M$_{\rm H_2}$~$=$~$\alpha L'_{CO}$ assuming a standard conversion factor of 
3.0$\times 10^{20}$ cm$^{-2}$ (K\,km\,s$^{-1}$)$^{-1}$.  This corresponds to an 
$\alpha$ of 4.8 which implies a molecular gas mass of $2.26 \pm 0.29 \times 10^{9}$~ 
M$_{\odot}$.
This gas mass is comparable to the gas mass
found in most other CO detected BCGs \citep{edg01,sc03}, some giant 
ellipticals such as 3C236 \citep{lab12}
and is consistent with the dynamical mass of the disc 
(M$_{dyn,2.2}$ = 6.0\,$\times$\,10$^{10}$\,M$_{\odot}$).

Finally there is significant radio continuum at both 3 and 1~mm detected
as an excess baseline in the EMIR observations (285$\pm$30 and 128$\pm$18~mJy at 3 and 1~mm, respectively). This
emission is consistent with recent high frequency observations of the
core and lobes with MUSTANG on the GBT \citep{cot09} and implies that the vast majority, if not all,
of the detected 850~$\mu$m SCUBA flux of 69.3~mJy \citep{zem07} 
is from the radio continuum and not from dust emission.

\subsection{Atomic Gas}
\begin{figure}
\psfig{figure=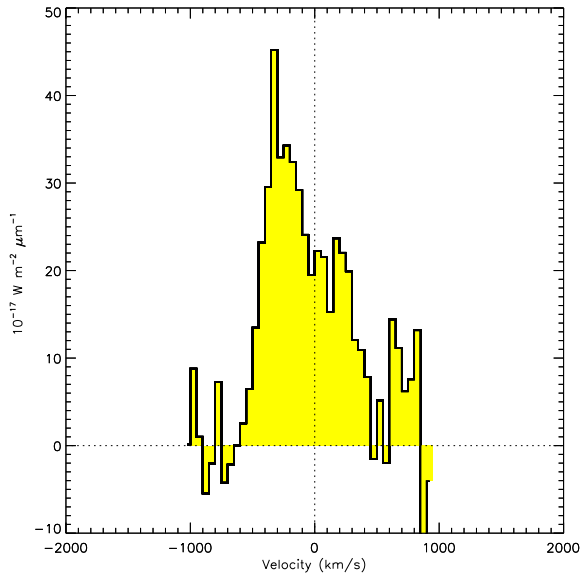,height=8cm}
\caption{Far infrared [OI] spectrum of Hydra-A taken with the PACS 
spectrometer on the {\sl Herschel} space telescope.  The width and structure of the line 
are consistent with the CO(2-1) although the ratio of the flux in the two peaks is 
different between the two lines.}
\label{fig:oi}
\end{figure}

We show the [OI] spectrum of Hydra-A taken with {\sl Herschel} in Figure \ref{fig:oi}.
The spectrum shows a striking similarity to the velocity structure of the CO(2-1) 
line implying that they trace the same cold gas clouds.  This correspondence between 
atomic and CO velocity profiles has also been observed in other BCGs at the centre 
of cooling flows \citep[for example A1068 and A2597,][]{edg10s,trm12s}.  We note however, 
that the [OI] line is substantially more asymmetric than the CO(2-1), we discuss this asymmetry 
in more detail in Section \ref{sec:bal}.
The intrinsic velocity 
resolution of the PACS spectrograph at the longest wavelengths makes it harder to 
match the velocity structure of the [CII] line \citep{edg10s}. However, if the 
CO(2-1) and [OI] lines are smoothed to the intrinsic resolution of PACS
at [CII] then the three lines are consistent.  

From these observations it is possible to estimate the mass of atomic gas present 
in the BCG. Following \citet{wol90}, and assuming the cooling rate and C$^+$ abundance 
determined therein, one can estimate the 
atomic gas mass as deduced from [CII] to be
\begin{equation}
M_{HI} = 2.7 \times 10^6\left(\frac{D}{\rm Mpc}\right)^{2}\left(\frac{F_{\rm [CII]}}{10^{-17} \, \rm W \, cm^{-2}}\right) M\odot
\label{eq:mhi}
\end{equation}
where $D$ is the distance to the cluster in Mpc, and F$_{\rm [CII]}$ is the flux of [CII] in 
W cm$^{-2}$.  We measure a [CII] flux of 5.3$\times$ 10$^{-21}$ W cm$^{-2}$ and use 
a distance of 242 Mpc,  which gives an atomic gas mass of 8.4$\times 10^{7}$ M$_{\odot}$. 
The assumptions going into this estimate make it rather uncertain (perhaps
by as much as an order of magnitude) but this mass is just 
below the upper limit of 4$\times 10^8$~M$_{\odot}$ derived from HI absorption by \citet{dwa94}.

\subsection{Star Formation and Dust}

  The dust lane visible in the HST imaging suggests that a substantial mass
  of dust must be present within the disc. Oonk et al. (in prep) calculate
  that the dust mass within the BCG to be 5.5 $\times 10^6$ M$_\odot$, at a
  temperature of 26~K, which corresponds to a gas to dust ratio of $\sim$ 400.
  However, the dust emission is unresolved with PACS at the shortest
  wavelength so the dust is not extended on scales of $>5''$.
  This gas to dust ratio is higher than in spiral galaxies with gas discs such as the 
  Milky Way or those in the SINGS survey (gas to dust $\sim$ 150 and $\sim$ 200 
  respectively, 
  \citet{dra07s}) suggesting the dust is more diluted in Hydra-A. However, 
  it is important to note the uncertainties on both the gas and dust mass when 
  making this comparison. The standard conversion factor from CO line intensity 
  to molecular gas mass assumes the gas has conditions (excitation, density, 
  temperature and metallicity) similar to molecular clouds in the Milky Way. There 
  is also some uncertainty in the dust mass given that its estimation is very 
  model dependant, and either of which could account for this difference. Therefore,
  a factor of two in the gas to dust ratio is not sufficient to draw any
  general conclusions about the nature of the dust in Hydra-A.
  
  The origin of the
  large mass of molecular gas within the system is best explained through
  the direct cooling of hot gas from the ICM as gas rich mergers are
  rare \citep{kirk09}
  and gas from stellar mass loss is most likely to be assimilated
  into the ICM \citep{mat90}, see Section \ref{sec:fule}.
  However, in this scenario the
  origin of the dust is not clear.  It is possible that the dust forms
  directly in the cold clouds deposited by the cooling flow
  \citep{fab94-2} or from stellar mass loss \citep{VD11}.  
  While both would produce less dust and therefore a higher gas to dust 
  ratio, the direct evidence for either is marginal \citep{raw12}. 
  Another possibility is that the dust is produced
  by the star formation within the disc but it requires a relatively
  fast pollution of the gas with dust to explain the observed gas to dust
  ratio.

  The Hydra-A BCG has a published star formation rate (SFR) of $\sim$2--3
  M$_{\odot}$yr$^{-1}$ from both UV and MIR observations
  \citep{hof12,don11}.  The UV image presented in Figure \ref{fig:fors}
  shows clear clumps of UV emission with a lack of UV from the region of
  the disc.  The concentration of dust in the gas disc could easily 
  obscure the UV emission from that region so it is likely star formation 
  is occurring in, and close to, the disc.

\section{Discussion}
\label{sec:dis}
\begin{figure}
\psfig{figure=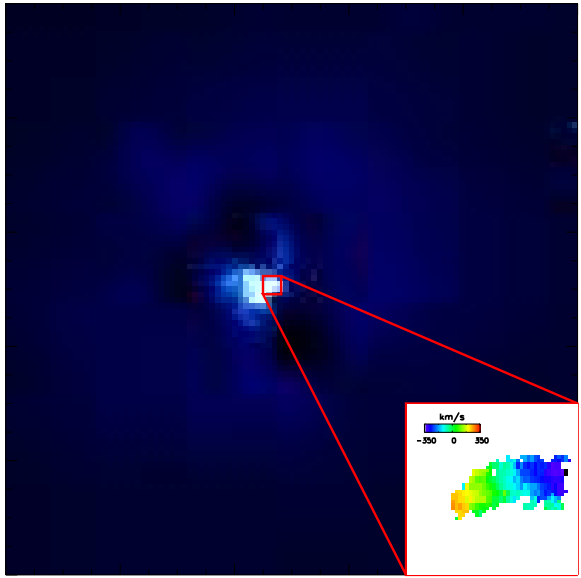,height=8cm}
\caption[Comparison of the Pa$\alpha$ disc to the position of X-ray cavities.]{The central $\sim$3x3 arcmin region of the Hydra A cluster shown in an unsharp masked 221 ksec 
integration X-ray image.  Two depressions in the X-ray emission show the location of 
cavities $\sim$20 arcsec to the north and south of the cluster centre.  The inset 
shows the Pa$\alpha$ velocity field in a $\sim$6x6 arcsec$^2$ region centred on 
the BCG as derived from single component fits to the Pa$\alpha$ line in Hydra A.  
The velocity field is with respect to the zero-point of the nominal cluster redshift 
of $z=0.0549$. The velocity field shows an east-west alignment almost perpendicular 
to the alignment of the X-ray cavities.}
\label{fig:alignment}
\end{figure}

The results presented above build directly on the previous studies of
Hydra-A and confirm unambiguously that there is a coherent and
large amplitude velocity gradient ($\sim700$~km~s$^{-1}$) in the gaseous disc across 
the galaxy.
Importantly, the major axis of this velocity gradient is aligned perpendicular to the 
axis of the cavities formed by the radio jets (Figures \ref{fig:fors} \& \ref{fig:alignment}), similar to 
the alignment seen in other BCGs \citep{wil09,oon10}. 
This strongly suggests that the angular momentum of the gas on large scales
influences the flow of gas close to the black hole and the generation
of jets \citep{baum92,pran10}.
The addition of the multiple molecular and atomic line detections
allows us to connect the cold gas to the observed emission line dynamics 
from IFU observations. The double-peaked velocity profile
seen in both the CO and [OI] lines is consistent with the cold gas
being distributed throughout the disc. This match extends to
both the ionised hydrogen (traced by Pa$\alpha$)
and [OI] gas sharing a significant asymmetry to the blue
that appears to be localised to the western edge of the gas disc.

The prominence of the gas disc in Hydra-A is accentuated by
the fact that it is being viewed close to edge-on and exhibits 
a striking dust lane in HST (Figure 1, Mittal, et al., in preparation) and ground-based imaging
\citep{ram11}. Similar gas discs are known
in other brightest cluster galaxies at the centre of cooling flows, e.g. NGC1275/3C84 in the
Perseus cluster \citep{wil05,lim08}
and NGC383/3C31 \citep{lim00,oku05,oca10}.
The prevalence of these discs in radio luminous systems is
suggestive of a link between large scale gas discs and AGN activity 
but may be an observational selection effect as
many of the galaxies with known discs were selected for HST imaging on
the basis of their radio power \citep{dek00,ver05,trm07}. 

The gas distribution within the disc as a function of radius cannot
be unambiguously determined from our current data so the innermost
extent of the gas and whether it is continuous with radius are
open issues. Only high spatial resolution imaging of optically
thin lines that are unaffected by obscuration will be able to
address the nature of the gas close to the Bondi radius and
the inner stable orbit of the gas \citep{pran10}.

\subsection{Fuelling the feedback?}
\label{sec:fule}
Hydra-A offers several significant insights into the
nature of cooling flows and AGN feedback.
The discovery of a distinct disc of cold gas in such
a powerful radio galaxy can be used to assess the 
energetics and fuelling of AGN activity.

Studies of the cavities in the intracluster medium in several 
cluster cores indicate that
a substantial amount of energy is being transferred from the AGN into
the surrounding gas through the mechanical work done by the radio lobes
\citep{mcn01,mcn07}. The total energy
of the outbursts is of the order $10^{61}$~erg which corresponds to the 
accretion of 10$^{7}$~M$_\odot$ of gas (at a rate of 0.1--0.25 M$\odot$ yr$^{-1}$) 
on to the central
super massive black hole assuming an efficiency of 0.1 \citep{wis07}. 
Combining our estimates of the gas masses in the various phases 
we calculate the total mass of cold gas in the disc to be 
2.3 $\pm$ 0.3 $\times$ 10$^9$~M$_{\odot}$.  Oonk et al. (in prep) calculate 
that the dust mass is 5.5 $\times 10^6$~M$_{\odot}$ at a temperature 26~K, though it is important 
to note that dust mass estimates are highly model dependant.
Assuming that the dust lane in the disc contains the majority of 
this cold material, then it has a total mass of 
2.3 $\pm$ 0.3 $\times$ 10$^9$ M$_{\odot}$, which is sufficient
to fuel a future outburst of comparable magnitude. We note however, that 
the infall time for gas in the disc will be large and that any gap within the 
inner disc could inhibit gas accretion. 

While this comparison is by no means proof that the cold
gas was solely responsible for the fuelling of any outburst
in Hydra-A, other systems may provide clues.
\citet{edg01} note that the BCGs with
the most powerful extended radio emission (Cygnus-A, Perseus-A,
M87 and Hercules-A) are less luminous in CO for their
given optical line luminosity. If the radio emission 
is generated by the rapid depletion of the gas in the core
of the galaxy then this could explain the weaker
molecular gas emission.

The origin of the cold gas is another open question. While the
accretion of cold gas from infalling cluster members and through
stellar winds can be invoked for lower luminosity systems
like M87 and NGC4696 \citep{spa89,spa04}, these stochastic
and ad hoc processes struggle to
explain the observation that
the vast majority of BCGs in cooling flows exhibit optical line emission
\citep{cra99,cav08} and CO lines \citep{edg01,sc03}
when the central cooling time or entropy is low, particularly given the
large gas masses ($>10^{10.5}$~M$_\odot$) found in many
systems. Instead, X-ray observations allow for residual gas cooling that is not
truncated by the AGN feedback which can still be significant \citep{wis07,sf11}
and the 10--100~M$_\odot$~yr$^{-1}$ observed
in the strongest cooling flows can result in the accumulation of
sufficient cold gas in $\sim10^9$~yr. This reasonably constant
supply of cold gas could in principle power the cores of all
BCGs in the strongest cooling flows all the time. The observation
that all strong cooling flows have a radio source \citep{mit09}
may be an indication that this consistent accumulation of fuel for powering the 
black hole and star formation is indeed the case.

If the accretion of cold gas deposited from the cooling of intracluster
gas is the dominant accretion mode in cluster cores, then what properties
do we expect other systems to share with Hydra-A? The very large
dust disc and the extremely broad line width in Hydra-A may
be the product of a preferential viewing angle where the
cold disc of gas and dust is being observed almost edge-on  (Figure \ref{fig:fors}).
This fortuitous viewing angle also maximises the visibility of the X-ray cavities if they
are created perpendicular to the disc. NGC1275 does not show
a similar dust lane but the presence of an inner spiral
structure and a direct line of sight to the nuclear
emission implies that we are viewing any disc at close to face on ($<30^\circ$). 

Finally, the observation of a disc implies that the cold gas has 
significant angular momentum with respect to the central galaxy. 
If the central galaxy is at rest in the cluster and any gas cooling
is symmetric then the angular momentum of the cold gas should be small.
However, optical observations of clusters show that the central galaxy is
frequently offset from the cluster mean \citep{zab93,bird94}
and X-ray observations imply that large scale bulk motion,
or `sloshing', is common in the ICM \citep{mar07,zuh10}.
Therefore, the presence of angular momentum in the cold gas phase is
not a surprise on scales beyond the Bondi radius. However, this
angular momentum poses a `centrifugal barrier' to gas accretion within it
\citep{nar11,mn12} so constraining how the gas moves within the disc
is an important observational challenge.


\subsection{The blue shifted line emission}
\label{sec:cf}

The extended, one-sided nature of the blue shifted Pa$\alpha$ line
component (see region 2 in Figure \ref{fig:pa}) implies that this emission is 
from gas that is not directly related to the gas disc
or the disc is asymmetric.
In NGC1275 gas on 5--20~kpc scales is observed which
does not follow the rotation observed on smaller scales
\citep{sal08,lim08,mit12} so the
presence of gas that has a substantial offset from the
systemic velocity of the BCG is not unprecedented. Indeed,
there are systems where the majority of the optical
line emission lies off the BCG spatially and in velocity,
such as A1991 and Ophiuchus \citep{ham12}. In these
cases the peak of the X-ray emission is coincident with
the line emission and not on the BCG. In the case of
Hydra-A, the X-ray emission close to the BCG is 
strongly affected by the radio cavities but there is 
a relatively weak excess to the west of the BCG \citep{sam00,wis07}.

What are the possible origins of this additional gas component?
One possibility is that it is related to the
merger of another gas-rich galaxy with the BCG.
However, we view this scenario as unlikely  given
the relatively large gas mass involved and the lack of
any donor galaxy in the cluster core. 
The inference
that the cold gas found in the cores of most clusters
is related to gas cooling rather than a merger with
a gas-rich galaxy has been the subject of extensive
debate over the past three decades \citep{spa92,edg01}
but the close connection between the X-ray properties
of the cluster core and the optical, infrared and radio 
properties of the BCG which is seen in the vast majority
of systems clearly favours a sustained
and non-stochastic explanation \citep{cav08,voi08,ode08,san09}. 
Therefore, the origin of the cold gas from the
direct cooling of the ICM best matches the 
observations.

One implication of the gas cooling from the ICM is that
it will retain the velocity of the cluster and not 
necessarily that of the BCG. So,
if the BCG is not at rest with 
respect to the cluster, then in principle it is possible
that two velocity components could be observed,
one associated with the cluster and one with the BCG.
This could potentially explain the second component seen in 
the Pa$\alpha$. 
To test whether this is the case in Hydra-A,
we extracted the galaxies known within 15$'$ (940~kpc)
from NED and find 20 potential cluster members (other than
Hydra-A itself) and the mean redshift for these is 0.0529
or 450 km~s$^{-1}$ lower than Hydra-A and
comparable to the blue-shifted line component.
Some caution is required in interpreting this result
given the number of galaxies is relatively small and that
there is a second cluster of galaxies present (probably
associated with Abell 780 that lies 11$'$ from Hydra-A
and is frequently, and completely erroneously, equated to Hydra-A in the literature).
However, there is some evidence that the second gas component
could be related to the cluster velocity giving a natural
explanation as to why two velocity components are present.

This conclusion requires the BCG to have a significant
velocity with respect to the cluster. There is a considerable
literature on ``speeding cDs'' \citep{bg83,zab93,pim06}
and there is a significant fraction of clusters with a velocity
offset between the cluster mean and the BCG velocities of more than
300~km~s$^{-1}$. The origin of this velocity  offset is poorly
understood but probably relates to a recent merger of a substantial
subcluster that affects the cluster mean velocity with respect to 
the BCG \citep{fuj06}. The presence of cold fronts and merger shocks could
be used to support this interpretation \citep{mv07,zuh10,zuh11}.
However, these are most prominent for mergers in the plane of the sky,
whereas the observed velocity difference will be largest for
mergers along the line of sight so the lack of these features
does not exclude the presence of a significant merger. 

The implications for the interpretation of other clusters
are significant. For instance, A1795 has a broad CO
velocity width \citep{sc03} and a prominent
tail of H$\alpha$ line emission \citep{ode04,cra05,mcv09}.
The BCG in A1795 is known to be offset from the cluster mean \citep{oeg94} and there
is evidence for a merger from the X-ray imaging \citep{ett02}. Also
the velocity offsets between the stellar component of the BCG,
the cold gas retained by the BCG, the mean of the cluster members
and the cold gas outside the influence of the BCG all need to
be considered separately. The gaseous nature of the ICM and cold
molecular gas mean that they can experience shocks and 
decouple from the dark matter
and stellar components which are relatively collisionless and 
respond only to gravity. Therefore,
it is possible that all four components could differ in velocity, with
the BCG and cluster mean having the widest separation and the 
gas components sitting between them. Accurate measurements
of the stellar and ionised gas components of BCGs are relatively
straightforward to obtain \citep[e.g.][]{cra99} as are similar 
measures of the molecular and cool atomic gas. 
However, determining the mean velocity 
of cluster members requires at least 50 cluster members to
ensure a statistically robust value so dedicated multi-object
spectroscopy is required which has more demanding observing time requirements.

One concern with this interpretation is that gas kinematics in this region 
remain consistent with the rotating disc rather than appearing as a separate 
kinematic component.  The other possible explanation of the second peak in the 
Pa$\alpha$ flux map is that there is some asymmetry in the disc. We note that 
the H$_2$ (1--0) S series lines do not detect this second component at greater 
then 5$\sigma$ significance (see Figure \ref{fig:s0}), suggesting the gas may be 
more ionised 
at the western edge of the disc.  The presence of a bright region of UV 
emission near to this second peak (see region 2 in the left hand panel of 
Figure \ref{fig:fors}) could potentially account for a higher fraction of ionised
gas in this region (see Section \ref{sec:bal} for discussion on the balance of gas 
phases).  Another possible explanation for an asymmetry in the disc is that it is the result of 
gas which has been lifted out of the central regions.  The excavation of material
from BCGs by radio jets has been seen in several systems (e.g. Perseus \citep{fab03a} and 
Abell 2052 \citep{bla01}) and if occurring on small scales here could produce an asymmetry 
in the disc on the scales seen. While the radio map shown in Figure \ref{fig:fors} 
\citep{tay90,sam00} does show an extent to the northwest which matches the position and 
direction of blue shifted emission \citet{tay90} identify this as an artificial 
feature produced by phase errors. 

\begin{figure}
\psfig{figure=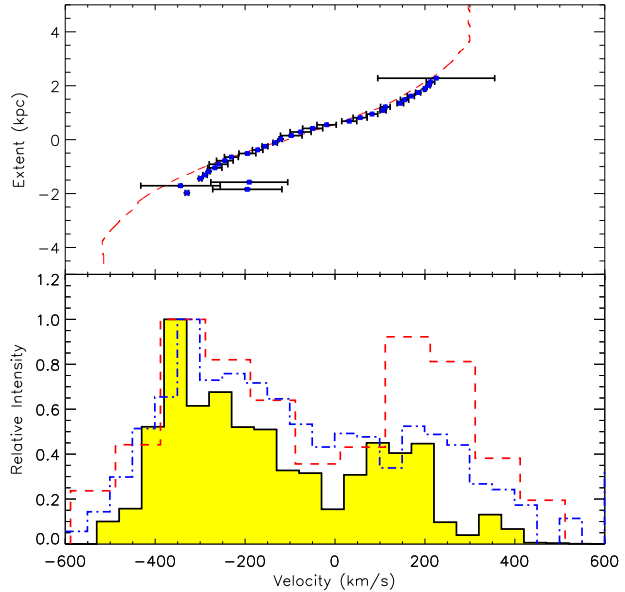,width=\linewidth}
\caption{Here we show the Pa$\alpha$ velocity profile along with a 
reconstructed spectrum.  The reconstructed spectrum of the 
Pa$\alpha$ bears a remarkable resemblance to those of the atomic (blue, 
dot-dashed line) and molecular (red, dashed line) line profiles.
The Pa$\alpha$ profile shows an asymmetry with more pronounced blue peak 
similar to that of the [OI] suggesting they trace similar gas distributions.
Note that all three show a double peaked structure with a similar separation 
indicating that the CO and [OI] share the same velocity structure as the 
Pa$\alpha$.}
\label{fig:sb}
\end{figure}

\subsection{The balance of molecular, atomic and ionised gas}
\label{sec:bal}
The SINFONI data provide a direct comparison of the molecular
and ionised gas in Hydra-A from the Pa$\alpha$ and
(1-0) S Series H$_2$ molecular lines.
The presence of a secondary peak in the Pa$\alpha$ which is not present in the 
warm molecular gas at above 5$\sigma$ suggests the western edge of the disc is either 
less molecular or cooler than the rest of the disc.


In order to compare the cold molecular, atomic and ionised gas directly we produced a 
reconstructed spectrum of the Pa$\alpha$ by taking the measured velocity, 
linewidth and flux of the emission at each point used to produce the velocity profile 
and created a Gaussian with 
these parameters. The spectra of these individual points were then summed to give 
the final spectrum of the Pa$\alpha$ emission before being binned to a level comparable 
to the CO and [OI] spectra (50 km s$^{-1}$). As this reconstructed spectrum 
is produced from the emission detected by the fitting routine it only 
includes the high surface brightness ($\geq$ 7$\sigma$ significance) Pa$\alpha$ emission.
A comparison of the CO, [OI] and reconstructed Pa$\alpha$ integrated line velocity 
profiles is shown in Figure \ref{fig:sb}. 
We note that the Pa$\alpha$ and [OI] line profiles are 
substantially more asymmetric than the line profile of the CO.  For the Pa$\alpha$ we 
attribute this to the secondary flux peak towards the western (blue shifted) edge 
of the disc, possibly related to the cooling flow (see Section \ref{sec:cf}).  
The [OI] profile shares a similar asymmetry to the Pa$\alpha$ 
suggesting that they are emitted by gas clouds sharing the same spatial distribution.
The more symmetric appearance of the CO profile suggests that it lacks a significant extra component, 
much as the (1-0) S Series H$_2$ molecular lines do.

However, despite this it can be seen that the three profiles are 
remarkably consistent with each other. All three profiles show a double 
peaked profile with a consistent separation between the peaks.  We
note that the Pa$\alpha$ line appears narrower then the CO and [OI]. While 
this difference may be real, it may also be an effect of the different 
spatial scales from which these spectra were extracted.  Both the CO and 
[OI] profiles are extracted from regions on the order of 10$\times$10 
arcsec$^2$ ($\sim$ 10 arcsec beam for the CO and the [OI] pixels scale is 
$\sim$ 10 arcsec). By contrast the Pa$\alpha$ profile is reconstructed from 
the 7$\sigma$ fits (a region of $\sim$ 5$\times$2 arcsec$^2$) and thus has a 
sharp cut-off at low surface brightness.

 Therefore, we infer that the gas on the western edge of
the disc does contain cold molecular gas implying that the non-detection of the 
warm, vibrationally excited molecular line emission is related to the excitation of
the gas not the absence of molecular gas. 
This difference in excitation could be due to lower shock
heating \citep{jaf01,wil02}, less direct, radiative AGN heating away from the centre of the galaxy
or the gas is more directly related to the star formation in the nearby UV bright region  
(see Figure \ref{fig:fors}). 
However, equally large differences are seen in the ratio
of Pa$\alpha$ and the molecular lines in IFU observations
of other systems with stronger cold molecular emission
away from the central galaxy \citep{wil09,oon10}
and the ratio varies significantly between sources
\citep{edg02} so one cannot draw universal conclusions
from the line ratios in Hydra-A alone.


\section{Summary}
\label{sec:sum5}
Mechanical feedback provided by AGN outbursts from BCGs is commonly believed to be 
the main contributing factor which prevents the catastrophic cooling of gas within 
cluster cores.  For this process to work effectively a mechanism of self regulation 
is required to maintain the balance between the cooling gas and the feedback.  This 
requires a direct link between the cooling gas (on kpc scales) and 
its supply to the super massive black hole be 
established which results in the activation of the AGN once cooling reaches a critical 
threshold.
Here we study Hydra-A, a well known BCG in which AGN feedback appears to be currently 
active. We present clear evidence for a kpc scale rotating disc of ionised and warm 
molecular gas. In addition, the agreement between the integrated line profile of
the cold molecular and atomic lines and the ionised
and warm molecular gas traced in the optical and NIR IFU
observations (Fig \ref{fig:sb}) implies that the cold gas follows a similar 
ordered rotation.  The axis of this rotation appears parallel to the axis along 
which the AGN is inflating cavities within the cluster core.

 
It is important to remember that for a single object this alignment could 
simply be coincidental so a larger more statistically representative sample is 
needed. However, it is suggestive given the presence of similar CO profiles and 
known molecular discs in other powerful radio galaxies (see Section \ref{sec:co}).  
It hints at a possible connection between the angular momentum of gas cooling 
on kpc--scales and the formation of the radio jets in the AGN.  
The existence of a relatively thin
($<$1~kpc) disc that has a velocity range of $>$700~km~s$^{-1}$
has profound implications for our understanding of Hydra-A
in particular and radio galaxies in general.


\section*{Acknowledgements}

SLH acknowledges the support received from an STFC 
studentship and the European Research Council for Advanced Grant Program num 
267399-Momentum. ACE acknowledges support from STFC grant ST/I001573/1. 
AMS gratefully acknowledges an STFC Advanced Fellowship through grant 
ST/H005234/1.

Based on observations made with ESO Telescopes at the La Silla or Paranal Observatories under programme ID 080.A-0224 and 082.B-0671

Herschel is an ESA space observatory with science instruments provided by European-led Principal Investigator consortia and with important participation from NASA.

Based on observations made with the NASA/ESA Hubble Space Telescope, obtained from the data archive at the Space Telescope Institute. STScI is operated by the association of Universities for Research in Astronomy, Inc. under the NASA contract NAS 5-26555.

Based on observations carried out with the IRAM 30m Telescope. IRAM is supported by INSU/CNRS (France), MPG (Germany) and IGN (Spain)

The National Radio Astronomy Observatory is a facility of the National Science Foundation operated under cooperative agreement by Associated Universities, Inc.

\bibliographystyle{mn2e}
\bibliography{bib}




\label{lastpage}

\end{document}